\documentclass[11pt,a4paper]{article}
\usepackage{jheppub}
\usepackage{epsf}
\usepackage{amsmath}
\usepackage{amssymb}
\usepackage{amsfonts}
\usepackage{graphicx}
\usepackage{graphics}
\usepackage{epstopdf}
\usepackage{url}

\newcommand{\nn}{\nonumber} 
\newcommand{\beq}{\begin{equation}}
\newcommand{\eeq}{\end{equation}} 
\newcommand{\beqa}{\begin{eqnarray}} 
\newcommand{\eeqa}{\end{eqnarray}}

\def\bi{\begin{itemize}}
\def\ei{\end{itemize}}
\def\be{\begin{equation}}
 \def\ee{\end{equation}}
\def\ben{\begin{equation*}}
 \def\een{\end{equation*}}
 \def\bea{\begin{eqnarray}}
 \def\eea{\end{eqnarray}}
 \def\bean{\begin{eqnarray*}}
 \def\eean{\end{eqnarray*}}
\newcommand{\ie}{{\it i.e.}}  \newcommand{\eg}{{\it e.g.}}

\newcommand{\RpA}{R_{\rm pA}}

\newcommand{\lqcd}{\Lambda_{_{\rm QCD}}}

\newcommand{\eq}[1]{(\ref{#1})}
\newcommand{\ed}{\end{document}}

\newcommand{\ave}[1]{\langle{#1}\rangle}

\newcommand{\mT}{M_\perp}

\newcommand{\jpsi}{{\mathrm J}/\psi}
\newcommand{\xf}{x_{\mathrm{F}}}

\newcommand{\pt}{p_{_\perp}}

\newcommand{\pp}{p--p}
\newcommand{\pA}{p--A}
\newcommand{\dA}{d--A}

\newcommand{\pN}{p--N}
\newcommand{\hi}{A--A}
\newcommand{\dd}{{\rm d}}
\newcommand{\half}{\frac{1}{2}}  
 \newcommand{\gsim}{\gtrsim}

\newcommand{\qzero}{\hat{q}_0}

\newcommand{\gevsqfm}{GeV$^2$/fm}

\newcommand{\ndf}{{\rm ndf}}

 \def\p{\vec{p}_\perp} \def\pt{p_{\perp}}

\def\b{{\rm \bf b}} \def\bt{\rm b}
\def\r{{\rm \bf r}} 
\def\s{{\rm \bf s}} 

\def\bm#1{\mbox{\boldmath$#1$}}
 
 \def\esim{\,\mathrel{\rlap{\lower0.2em\hbox{$-$}}\raise0.15em\hbox{\footnotesize $\hskip0.04em\sim$}}\,}
 \def\gsim{\mathrel{\rlap{\lower0.2em\hbox{$\sim$}}\raise0.2em\hbox{$>$}}}
 \def\ksim{\mathrel{\rlap{\lower0.2em\hbox{$\sim$}}\raise0.2em\hbox{$<$}}}

\def\xf{x_{_F}}

\newcommand{\sqrts}{\sqrt{s}}

\title{Centrality and ${\bm \pt}$ dependence of ${\bm \jpsi}$ suppression\\ in proton--nucleus collisions 
from parton energy loss}

\author[a,b]{Fran\c{c}ois Arleo,}
\author[c,d]{Rodion Kolevatov,}
\author[c]{St\'ephane Peign\'e,}
\author[c]{Maryam Rustamova}

\affiliation[a]{Laboratoire Leprince-Ringuet (LLR), \'Ecole polytechnique, CNRS/IN2P3 91128 Palaiseau, France}
\affiliation[b]{Laboratoire d'Annecy-le-Vieux de Physique Th\'eorique (LAPTh)\\ UMR5108, Universit\'e de Savoie, CNRS, BP 110, 74941 Annecy-le-Vieux cedex, France}
\affiliation[c]{SUBATECH, UMR 6457, Universit\'e de Nantes, Ecole des Mines de Nantes, IN2P3/CNRS \\ 4 rue Alfred Kastler, 44307 Nantes cedex 3, France}
\affiliation[d]{Department of High Energy Physics, Saint-Petersburg State University\\ Ulyanovskaya 1, 198504, Saint-Petersburg, Russia}

\emailAdd{arleo@lapth.cnrs.fr}
\emailAdd{kolevato@subatech.in2p3.fr}
\emailAdd{peigne@subatech.in2p3.fr}
\emailAdd{rustamov@subatech.in2p3.fr}

\abstract{The effects of parton energy loss and $\pt$-broadening in cold nuclear matter on the $\pt$ and centrality dependence, at various rapidities, of $\jpsi$ suppression in \pA\ collisions are investigated. Calculations are systematically compared to E866 and PHENIX measurements. The very good agreement between the data and the theoretical expectations further supports $\pt$-broadening and the associated medium-induced parton energy loss as  
dominant effects in $\jpsi$ suppression in high-energy \pA\ collisions. Predictions for $\jpsi$ (and $\Upsilon$) suppression in p--Pb collisions at the LHC are given.}

\keywords{Parton energy loss; heavy-quarkonium; cold QCD matter; proton--nucleus}
\arxivnumber{xxxx.yyyy}

\begin{document} 

\maketitle
\setcounter{footnote}{0}
\renewcommand{\thefootnote}{\arabic{footnote}} 	


\section{Introduction}

A wide range of phenomena observed in heavy-ion collisions at RHIC and LHC, both in hard and soft processes, suggests that a new, strongly interacting state-of-matter has been created, the quark-gluon plasma (QGP). One of the most striking observables is the suppression of high-$\pt$ particles (``jet-quenching'') in \hi\ compared to a simple scaling of \pp\ collisions~\cite{Adler:2003qi,Adams:2003kv,Aamodt:2010jd,CMS:2012aa}, which together with dijet momentum imbalance~\cite{Aad:2010bu,Chatrchyan:2012nia} finds a natural explanation in parton energy loss in the QGP~\cite{dEnterria:2009am}. Another observable in heavy-ion collisions which received much attention is heavy-quarkonium suppression. Such a suppression is expected from the Debye screening of the in-medium heavy-quark potential, and was thus originally proposed as a potential signal of QGP formation (and a direct probe of the plasma temperature)~\cite{Matsui:1986dk}. However it was later realized that several other effects can modify quarkonium yields in \hi\ collisions (see~\cite{Frawley:2008kk} for a review), some of those effects playing a role even in absence of any hot medium.

In order to quantify the properties of the QGP created in heavy-ion collisions -- a main goal of the forthcoming measurements at RHIC-II and LHC --  a solid understanding of the nuclear modification of particle spectra in cold nuclear matter is thus required. A baseline for the study is provided by \pA\ (or \dA) collisions where a significant suppression is already reported for some particle species. In particular, light hadron~\cite{Arsene:2004ux, Adler:2004eh} and $\jpsi$~\cite{Leitch:1999ea,Adare:2012qf} production in \pA\ collisions at forward rapidity is significantly lower than expected from the naive scaling of \pp\ spectra. 

At the same time it is astonishing that no consensus on the cold nuclear matter effects responsible for $\jpsi$ suppression has been achieved yet. 
In addition to the modifications of nuclear parton distribution functions,
various mechanisms have been proposed to explain $\jpsi$ suppression in p--A collisions. In the nucleus rest frame, a high-energy $\jpsi$ is formed long after the nucleus thus what actually propagates through the nucleus is the parent $c \bar c$ pair. Some approaches attribute $\jpsi$ suppression to an effective absorption cross section $\sigma_{\rm abs}$ of the $c \bar c$ pair (see Refs.~\cite{Kopeliovich:2011zz} and \cite{Ferreiro:2012sy} for recent works). 
Other models attribute $\jpsi$ suppression to the increase of the $c \bar c$ pair invariant mass by the multiple soft rescatterings through the nucleus, leading to a reduction of the overlap with the $\jpsi$ wave function \cite{Benesh:1994du} (see also \cite{Fujii:2006ab}). In the approach used by~\cite{Sharma:2012dy} and that of our group~\cite{Arleo:2012rs}, the dominant role is played by parton radiative energy loss, \ie, gluon radiation induced by multiple scattering of fast partons (or color octet $c \bar c$ pair) travelling through the nucleus. 
 
The present study, together with our previous works \cite{Arleo:2012rs,Arleo:2012hn,Arleo:2010rb}, supports parton energy loss induced by $\pt$-broadening as the main effect in $\jpsi$ suppression. At this point we should stress that the physical content of our approach and that of  \cite{Sharma:2012dy}, both based on parton energy loss, are actually quite different. 
In~\cite{Sharma:2012dy}, it is assumed that
the interference between gluon emissions off the incoming and outgoing parton participating to the hard production process can be neglected. Under this assumption the induced energy loss of an incoming gluon in $J/\psi$ production is parametrically the same (up to color factors) 
as that of the incoming quark in the case of Drell-Yan pair production~\cite{Neufeld:2010dz}.

It was however argued in~\cite{Arleo:2010rb} that when the outgoing parton (or compact $c\bar c$ pair in the case of $\jpsi$ production) is produced at small angle (\ie, large energy $E$ at limited $\pt$) in the target nucleus rest frame, the medium-induced gluon spectrum is dominated by the interference between initial and final state radiation of gluons with large formation times. Such radiation 
is expected in low-$\pt$ $\jpsi$ production in \pA\ collisions, where an incoming gluon is scattered at small angle into a compact color octet $c \bar{c}$ pair. The associated energy loss is proportional to the $\jpsi$ energy $E$ and the role of this effect is thus expected to increase with increasing rapidity. This observation is at the basis of the energy loss scenario used in~\cite{Arleo:2012rs} to describe $\jpsi$ suppression as a function of $y$ (or $\xf$) in a wide collision energy range for minimum bias \pA\ and \dA\ collisions. In the present study we generalize this approach to address the $\pt$ and centrality dependence of nuclear suppression at 
fixed-target (E866), RHIC and LHC energies. 

The paper is organized as follows. In Section~\ref{sec:model} we generalize the calculation scheme of~\cite{Arleo:2012rs} to address the $\pt$ and centrality dependence of $\jpsi$ nuclear modification factors. Section~\ref{sec:comparison} is devoted to the comparison of the model to E866 and latest PHENIX data
and predictions in p--Pb collisions at the LHC are given in Section~\ref{sec:LHCpredictions}. Results are summarized in Section~\ref{sec:conclusion}.

\section{Model}
\label{sec:model}

\subsection{Shift in $E$ and $\vec{p}_\perp$} 
\label{sec:shift}

Heavy-quarkonium\footnote{denoted by ``$\psi$'' in the rest of the paper.} nuclear suppression in minimum bias \pA\ collisions as compared to \pp\ collisions can be represented in terms of the ratio
\be
\label{RpA}
R_{\mathrm{pA}}^{\psi}(y, \pt) = \frac{1}{A} \, {\frac{\dd\sigma_{\mathrm{pA}}^{\psi}}{\dd y \, \dd^2 \p} \biggr/ \frac{\dd\sigma_{\mathrm{pp}}^{\psi}}{\dd y \, \dd^2 \p}}  \, ,
\ee
where $y$ and $\p$ are the quarkonium rapidity and transverse momentum in the c.m. frame of an elementary \pN\ collision (of energy $\sqrt{s}$). By convention forward (positive) rapidities correspond to the proton fragmentation region.

In the present study we generalize the model of Ref.~\cite{Arleo:2012rs} by expressing the quarkonium {\it double} differential cross section in \pA\ collisions in terms of that in \pp\ collisions, where a shift in the quarkonium energy $E$ (defined in the nucleus rest frame) as in Ref.~\cite{Arleo:2012rs} but also a shift in $\vec{p}_\perp$ account, respectively, for the energy loss $\varepsilon$ and transverse momentum broadening $\Delta \vec{p}_\perp$ of the octet $Q \bar{Q}$ pair propagating through the nucleus. The ``double shift'' in $E$ and $\p$ 
relates the  \pA\ and \pp\ double differential cross sections $\dd\sigma/\dd E \dd^2 \vec{p}_\perp$ as\footnote{The dependence of $\Delta \p$ on the azimuthal angle $\varphi$, $\Delta \p \equiv \Delta \p (\varphi)$, will be implicit in the following, and we will use the notations:
\be
\int_{\varphi} \equiv \int \frac{\dd\varphi}{2\pi} \ ; \ \ \int_{\varepsilon} \equiv \int \dd\varepsilon \ ; \ \ \int_{\p} \equiv \int \dd^2 \p \  {\rm etc.} \nn
\ee}
\be
\label{doubleshift-E}
\frac{1}{A}\frac{\dd\sigma_{\mathrm{pA}}^{\psi}}{\dd E \, \dd^2 \p} = \int_{\varphi} \int_{\varepsilon} \,{\cal P}(\varepsilon, E) \, \frac{\dd\sigma_{\mathrm{pp}}^{\psi}}{\dd E \, \dd^2 \p} \left( E+\varepsilon, \p -\Delta \p \right)\ .
\ee
The quantity ${\cal P}(\varepsilon, E)$ is the energy loss probability distribution or {\it quenching weight} (see Section~\ref{sec:quenching-weight}) associated to the medium-induced radiation spectrum in a target nucleus A as compared to a lighter (\eg, proton) target nucleus. The integral over $\varepsilon$ is bounded by $\varepsilon_{\rm max} = {\rm min}(E_{\mathrm{p}} -E, E)$, where $E_\mathrm{p} \simeq s/(2 m_\mathrm{p})$ is the projectile proton energy in the nucleus rest frame. (We work in the limit $\sqrt{s} \gg m_\mathrm{p}$, with $m_\mathrm{p}$ the proton mass.) For the time being we consider minimum bias \pA\ collisions, hence the normalization factor $1/A$ in the l.h.s. of \eq{doubleshift-E}, but the model will be generalized in Section~\ref{sec:bdep} to \pA\ and \dA\ collisions in a given centrality class. We will assume that $\Delta \p$ is uniformly distributed in the azimuthal angle $\varphi$, and has a modulus defined by \eq{broad-modulus}.

The relation between the \pA\ and \pp\ differential cross sections in $y$ and $\p$,
\be
\label{doubleshift-y}
\frac{1}{A}\frac{\dd\sigma_{\mathrm{pA}}^{\psi}}{\dd y \, \dd^2 \p} = \int_{\varphi} \int_{\varepsilon} \,{\cal P}(\varepsilon, E) \,\left[ \frac{E}{E+\varepsilon} \right] \, \frac{\dd\sigma_{\mathrm{pp}}^{\psi}}{\dd y \, \dd^2 \p} \left( E+\varepsilon, \p -\Delta \p \right) \, ,
\ee
can be simply obtained from \eq{doubleshift-E} by using
\be
y\left( E, \p \right) = \ln{\left( \frac{E}{E_{\mathrm{p}}} \, \frac{\sqrt{s}}{M_\perp} \right)} \, ,
\label{yofEpt} 
\ee
where $M_\perp=(M^2+\pt^2)^{\half}$, 
with $M$ the mass of the $Q \bar{Q}$ pair.  
In Eq.~\eq{doubleshift-y}, the variable $E$ is given by $E = E\left( y, \p \right) = E_\mathrm{p}\,e^{y} \, M_\perp / \sqrt{s}$, and the \pp\ cross section is evaluated at $y = y\left( E+\varepsilon, \p -\Delta \p  \right)$.

For the double differential \pp\ cross section $\dd\sigma_{\mathrm{pp}}/\dd y \dd^2 \p$, we will use a parametrization consistent with the \pp\ data, see Section~\ref{sec:ppfits}. In our model, the nuclear modification factor \eq{RpA} obtained from \eq{doubleshift-y} is thus fully determined by the quenching weight ${\cal P}(\varepsilon, E)$ and the transverse momentum broadening $\Delta \pt$.

\subsection{Quenching weight and $\Delta \pt$} 
\label{sec:quenching-weight}

In Ref.~\cite{Arleo:2012rs} the appropriate medium-induced gluon radiation spectrum ${\dd I} / {\dd \omega}$ in quarkonium production in \pA\ collisions (as compared to p--B collisions) was derived, as well as the associated quenching weight ${\cal P}(\varepsilon, E)$, 
\bea
\label{quenching}
{\cal P}(\varepsilon, E) = \frac{\dd I}{\dd\varepsilon} \, \exp \left\{ - \int_{\varepsilon}^{\infty} \dd\omega  \frac{\dd{I}}{\dd\omega} \right\}  \, , \hskip 3cm && \\
\label{spectrum}
\frac{\dd I}{\dd \omega} = \frac{N_c \, \alpha_s}{\omega \, \pi} \left\{ \ln{\left(1+\frac{\ell_{\perp {\rm A}}^2 E^2}{M_\perp^2 \omega^2}\right)} - \ln{\left(1+\frac{\Lambda_{\rm B}^2 E^2}{M_\perp^2 \omega^2}\right)} \right\} \, \Theta(\ell_{\perp {\rm A}}^2 - \Lambda_{\rm B}^2) \, , && 
\eea
where $\ell_{\perp {\rm A}}$ represents the accumulated transverse momentum transfer due to soft rescatterings in the target nucleus A, and $\Lambda_{\mathrm B} = {\rm max}(\lqcd,\ell_{\perp {\rm B}})$. 
The latter dependence on $\lqcd$ arises from vetoing induced gluon radiation with $k_\perp <  \lqcd$~\cite{Arleo:2012rs}.
Note that ${\cal P}(\varepsilon, E)$ is determined analytically in terms of the dilogarithm function ${\rm Li}_2(x)$,
\be
\label{quenching-dilog}
{\cal P}(\varepsilon, E) = \frac{\dd I}{\dd\varepsilon} \, \exp \left\{ - \frac{N_c \, \alpha_s}{2 \pi} \left[ {\rm Li}_2\left( - \frac{\Lambda_{\rm B}^2 E^2}{M_\perp^2 \varepsilon^2} \right) - {\rm Li}_2\left( - \frac{\ell_{\perp {\rm A}}^2 E^2}{M_\perp^2 \varepsilon^2} \right) \right] \right\} \, .
\ee

The {\it semi-hard} transfer $\ell_{\perp {\rm A}}$ is given by
\be
\label{lperpA}
\ell_{\perp {\rm A}}^2 = \hat{q}_{\rm A} \, L_{\rm A} \, ,
\ee
where $L_{\rm A}$ is the {\it effective} path length\footnote{For the present study we will use
$L_{\rm Be} = 3.24\,{\rm fm}$, $L_{\rm Fe} = 6.62\,{\rm fm}$,  $L_{\rm W} = 9.35\,{\rm fm}$, $L_{\rm Au} = 10.23\,{\rm fm}$ and $L_{\rm Pb} = 10.11\,{\rm fm}$ for minimum bias collisions, as well as $L_{\rm p}= 1.5\,{\rm fm}$~\cite{Arleo:2012rs}. The dependence of $L_{\rm Au}$ (resp. $L_{\rm Pb}$) on the centrality class in d--Au collisions at RHIC (resp. p--Pb collisions at LHC) is described in Section~\ref{sec:bdep}, see Table~\ref{tab:centr-Npart}, and in Appendix~\ref{app:bdep}.} across the target nucleus A and $\hat{q}_{\rm A}$ the transport coefficient~\cite{Arleo:2012rs}
\be
\label{qhat-model}
\hat{q}_{\rm A} = \hat{q}(x_{\rm A}) \equiv \hat{q}_0 \left( \frac{10^{-2}}{x_{\rm A}} \right)^{0.3}\ ;\ \ x_{\rm A} = {\rm min}(x_{0 {\rm A}}, x_2) \ ; \ \ x_{0 {\rm A}} \equiv \frac{1}{2 m_\mathrm{p} L_{\rm A}} \ ; \ \  x_2 = \frac{M_\perp}{\sqrt{s}} e^{-y} \, .
\ee
The parameter $\hat{q}_0 = 0.075 \, {\rm GeV}^2/{\rm fm}$ was extracted in Ref.~\cite{Arleo:2012rs} from a fit to the E866 data for $R_{\mathrm{W/Be}}^{\jpsi}$. We will use this value in the present study, which thus contains no free parameter.\footnote{Let us also note that in the fitting procedure of Ref.~\cite{Arleo:2012rs}, small values of $\hat{q}_0$ and thus of $\hat{q} \, L$ were explored, for which the dependence of \eq{spectrum} on $\lqcd$ is relevant. However, it turns out that the extracted $\hat{q}_0 = 0.075 \, {\rm GeV}^2/{\rm fm}$ is large enough to satisfy $\sqrt{\hat{q} \, L} > 0.25 \,{\rm GeV}$ for all values of $L_{\rm A}$ (including $L_{\rm p}$) and $x_2$ considered in the present study, as can be easily checked from \eq{qhat-model} using $L_{\rm p}= 1.5\,{\rm fm}$. The present study is thus independent of the value of $\lqcd$ provided $\lqcd \leq  0.25 \,{\rm GeV}$.\label{foot:lqcddep}} 

Finally, the transverse momentum broadening in \pA\ with respect to p--B collisions is simply given by 
\be
\label{broad-modulus}
(\Delta \p)^2 = \ell_{\perp {\rm A}}^2 - \ell_{\perp {\rm B}}^2 = \hat{q}_{\rm A} L_{\rm A} - \hat{q}_{\rm B} L_{\rm B} \, .
\ee
In our study we neglect the fluctuations of the broadening $\ell_{\perp}^2$ around the average value $\hat{q} L$. We have checked that assuming a distribution in $\ell_{\perp}$ of the Gaussian type, 
$P(\ell_{\perp}) \propto \exp{(-\ell_{\perp}^2 / \hat{q} L)}$, 
only slightly modifies (by at most $10\%$) the predictions for $R_{\mathrm{pA}}^{\psi}(y,\pt)$ presented in Sections~\ref{sec:comparison} and~\ref{sec:LHCpredictions}, without changing the overall shape of $R_{\mathrm{pA}}$. 

\subsection{Parametrization of the \pp\  cross section}
\label{sec:ppfits}

Similarly to Ref.~\cite{Arleo:2012rs}, we use for the double differential \pp\ cross section $\dd\sigma_{\mathrm{pp}}/\dd y \dd^2 \vec{p}_\perp$ a simple parametrization consistent with the available \pp\ data, rather than relying on some model-dependent quarkonium production mechanism in hadronic collisions. 

The double differential cross section of prompt $\jpsi$ and $\Upsilon$ production can be conveniently parametrized as 
\be
\frac{\dd\sigma_{\mathrm{pp}}^{\psi}}{\dd y\,\dd^2 \p}  = {\cal N} \times \left(\frac{p_0^2}{p_0^2+\pt^2}\right)^{m} \times \left(1- \frac{2 M_\perp}{\sqrt{s}} \cosh{y} \right)^{n} \equiv {\cal N} \times \mu(p_\perp) \times \nu(y, p_\perp) \ .
\label{pp-fit-2d}
\ee
At the LHC ($\sqrt{s}=7$~TeV) the values of the free parameters ${\cal N}$, $p_0$, $m$ and $n$ are obtained from a global fit of ALICE~\cite{Aamodt:2011gj}, ATLAS~\cite{Aad:2011sp} and LHCb~\cite{Aaij:2011jh} data on prompt $\jpsi$ production\footnote{The ALICE data are only given for \emph{inclusive} $\jpsi$ production, \ie\ including $\jpsi$ from $B$-decays.} 
and from a fit of LHCb data~\cite{LHCb:2012aa} on $\Upsilon$ production. They are summarized in Table~\ref{tab:param} together with the corresponding $\chi^2/\ndf$ values.\footnote{The value of ${\cal N}$ is irrelevant as we consider only cross section \emph{ratios} and is therefore not given.}

\begin{table}[h]
 \centering
 \begin{tabular}[c]{cccccc}
   \hline
   \hline
 Quarkonium   & $\sqrt{s}$ (GeV) & $p_0$ (GeV) & $n$ & $m$ & $\chi^2/\ndf$ \\
\hline
$\jpsi$ & $7000$ &  $4.2$ & $19.2$ & $3.5$ & $54/139$ \\
$\Upsilon$ & $7000$ &  $6.6$ & $13.8$ & $2.8$ & $48/72$ \\
\hline
$\jpsi$ & $200$ &  $3.3$ & $(8.3)$ & $4.3$ & $27/37$ \\
$\jpsi$ & $38.7$ &  $3.1$ & $(4.5)$ & $5.3$ & $45/19$ \\
\hline
\hline
\end{tabular}
 \caption{Values of the fit parameters obtained from a fit to LHC ($\jpsi$ and $\Upsilon$), PHENIX, and E789 p--p data. Values given in parenthesis are fixed in the fitting procedure (see text for details).}
 \label{tab:param}
 \end{table}

At RHIC the amount and precision of data are not sufficient to fix precisely the fit parameters. The fit to the double differential data measured by PHENIX~\cite{Adare:2006kf} is therefore performed by fixing the value of $n=8.3$ obtained from the fit of the single differential cross section $\dd\sigma_{\mathrm{pp}}^{\jpsi}\big/\dd {y}$, see Ref.~\cite{Arleo:2012rs}. 

Finally, a fit to the $\jpsi$  E789 ($\sqrts=38.7$~GeV) data~\cite{Schub:1995pu} is also performed with the aim to compare the model predictions to the E866 data~\cite{Leitch:1999ea} at the same center-of-mass 
energy.\footnote{Note however that the E789 data have been taken at $\xf=0$ only. Therefore this parametrization may not be valid at large $\xf$ for which the model is compared to the E866 measurements, see Section~\ref{sec:comparisone866}.}
Like at RHIC, the number of fit parameters is reduced by fixing the value of $n=4.5$ obtained from fitting $\dd\sigma_{\mathrm{pp}}^{\jpsi}\big/\dd {\xf}$ data at this energy~\cite{Arleo:2012rs}.

\begin{figure}[t]
\begin{center}
    \includegraphics[width=14cm]{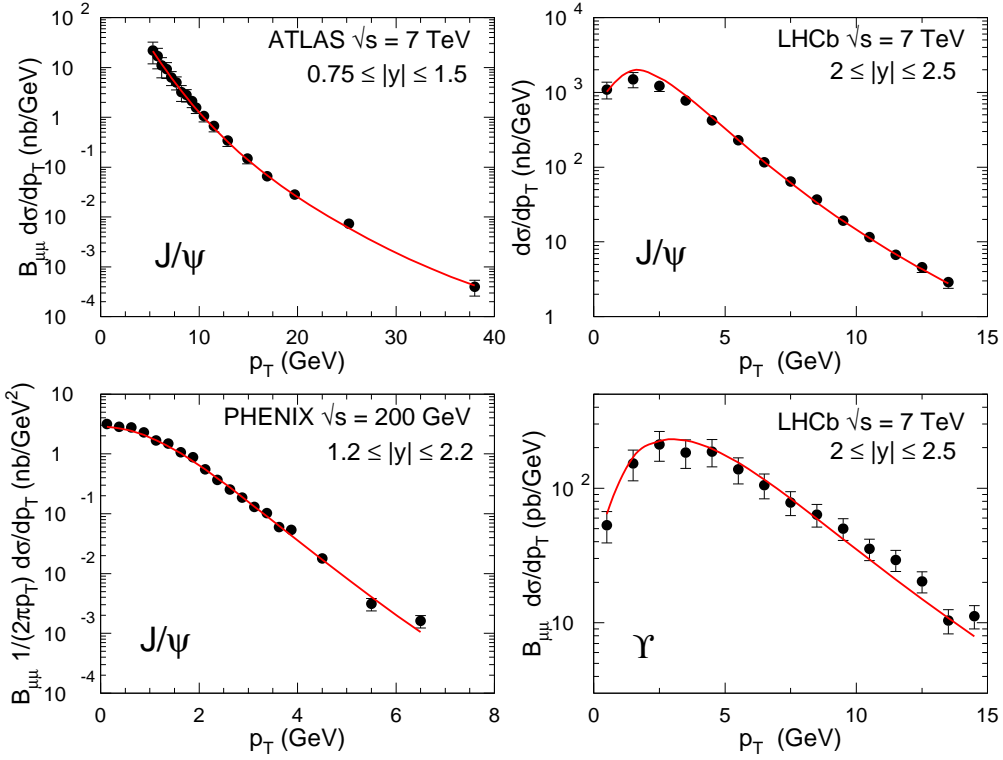}
  \end{center}
\vspace{-1cm}
\caption{Comparison between heavy-quarkonium ($\jpsi$, $\Upsilon$) production data in p--p collisions and the parametrization 
\eq{pp-fit-2d} (solid red line). Data are taken from ATLAS~\cite{Aad:2011sp}, LHCb~\cite{Aaij:2011jh,LHCb:2012aa}, and PHENIX~\cite{Adare:2006kf}.}
  \label{fig:fit2d}
\end{figure}

The agreement between some of the prompt $\jpsi$ and $\Upsilon$ measurements at RHIC and LHC and the parametrization 
\eq{pp-fit-2d} is shown in Fig.~\ref{fig:fit2d}. Finally, let us mention that the functional form of the parametrization \eq{pp-fit-2d} is fully consistent with that for the single differential rate in $\xf$ used in~\cite{Arleo:2012rs}, as we briefly show in Appendix~\ref{app:compare-param}.

\subsection{A useful approximation for $R_{\rm pA}^{\psi}(y, \pt)$}

Using \eq{doubleshift-y}  and \eq{pp-fit-2d} the attenuation factor \eq{RpA} reads
\bea
\label{rpa-fact}
R_{\rm pA}^{\psi}(y, \pt) &=& \int_{\varphi} \int_{\varepsilon}  \,{\cal P}(\varepsilon, E) \, \left[ \frac{E}{E+\varepsilon} \right] \, \frac{\mu(|\p -\Delta \p |)}{\mu(\pt)} \, \frac{\nu(E+\varepsilon,\p - \Delta \p)}{\nu(E,\pt)} \\ 
&=&  \int_{\varepsilon}  \,{\cal P}(\varepsilon, E) \, \left[ \frac{E}{E+\varepsilon} \right] \frac{\nu(E+\varepsilon,\pt)}{\nu(E,\pt)} \int_{\varphi} \frac{\mu(|\p -\Delta \p |)}{\mu(\pt)}\,\frac{\nu(E+\varepsilon,\p - \Delta \p)}{\nu(E+\varepsilon,\pt)} \, . \nn 
\eea
Since ${\cal P}(\varepsilon, E)$ is peaked at small values of $\varepsilon$, we neglect $\varepsilon$ in the latter $\varphi$ integral. In this approximation, the $\varphi$ and $\varepsilon$ integrals factorize,
\be
\label{rpa-fact-2}
R_{\rm pA}^{\psi}(y, \pt) \simeq R_{\rm pA}^{\rm broad}(y, \pt) \cdot R_{\rm pA}^{\rm loss}(y, \pt) \, , 
\ee
where
\bea
\label{eq:rpabroad}
R_{\rm pA}^{\rm broad}(y, \pt) &\equiv& \int_{\varphi} \frac{\mu( |\p -\Delta \p|)}{\mu(\pt)}\,\frac{\nu(E,\p - \Delta \p)}{\nu(E,\pt)}  \, , \\
\label{eq:rpaloss}
R_{\rm pA}^{\rm loss}(y, \pt) &\equiv& \int_{\varepsilon} \,{\cal P}(\varepsilon, E) \, \left[ \frac{E}{E+\varepsilon} \right] \, \frac{\nu(E+\varepsilon, \pt)}{\nu(E, \pt)}  \, .
\eea

The factor $R_{\rm pA}^{\rm broad}(y, \pt)$ describes nuclear modification due to transverse momentum broadening only, as can be seen by setting ${\cal P}(\varepsilon, E)= \delta(\varepsilon)$ in \eq{rpa-fact}. The factor $R_{\rm pA}^{\rm loss}(y, \pt)$ describes the effect of energy loss only, obtained by setting $\Delta \pt = 0$ in \eq{rpa-fact}. In the following we will use the factorized expression \eq{rpa-fact-2}, which turns out to be a very accurate approximation to \eq{RpA} 
in all the practical applications of Sections~\ref{sec:comparison} and~\ref{sec:LHCpredictions}.

Finally, let us mention that the $\pt$-inclusive nuclear suppression factor $R_{\rm pA}^{\psi}(y)$ studied in Ref.~\cite{Arleo:2012rs} can be simply recovered from \eq{rpa-fact}, along the same lines as in Appendix~\ref{app:compare-param} where the parametrization of  $\dd\sigma_{\mathrm{pp}}/\dd y$ is obtained from $\dd\sigma_{\mathrm{pp}}/\dd y \dd^2 \p$. Integrating both the numerator and denominator of \eq{RpA} (and thus of \eq{rpa-fact}) over $\p$, the function $\nu(y, \pt)$ can be replaced by its value at a typical $\bar{p}_\perp$ determined by the width of $\mu(\pt)$ (see \eq{medianpt}), and the $\p$-integral of $\mu(\pt)$ cancels between the numerator and denominator. As a result, the $\pt$-inclusive suppression factor
reads $R_{\rm pA}^{\psi}(y) \simeq R_{\rm pA}^{\rm loss}(y, \bar{p}_\perp)$, 
which corresponds exactly to the quantity studied in Ref.~\cite{Arleo:2012rs}. 

\subsection{Centrality dependence}
\label{sec:bdep}

In the preceding sections we addressed the case of minimum bias \pA\ collisions. The model is generalized to the case of \pA\ collisions at a given centrality 
(or at a given impact parameter $\bt$) 
by using the effective length $L_{\rm A}$ corresponding to that centrality. 
Since $L_{\rm A}$ fully determines the essential quantities $\ell_{\perp {\rm A}}$ and $\Delta \pt$ of the model (see \eq{lperpA}, \eq{qhat-model} and \eq{broad-modulus}), this is the only modification required. However, the impact parameter (as well as the number of participants) is not a direct experimental observable, and the consistent way for making a theoretical estimate of 
$L_{\rm A}$ is to follow the experimental procedure as closely as possible.

Both at RHIC~\cite{Adler:2007aa} and the LHC~\cite{Abelev:2012ola} a centrality selection is done by triggering on event multiplicity in forward detectors. This multiplicity is strongly correlated with the number of participating nucleons from the target nucleus. Thus, multiplicity cuts impose a restriction on the number of participants in a given event. The values of the forward event multiplicities which separate centrality classes in the experiment are chosen to attribute a certain fraction of the total inelastic cross section to each centrality class. The common choice is making 20\% slices from the most central (largest multiplicity) to the peripheral (lowest multiplicity) events. The exception is the most peripheral class which is taken at 60--88\% of the total inelastic cross section at RHIC (class D) and 60-100\% at the LHC (class 4).

To compute the average path length $L_{\rm A}$ for each centrality class, we employ the following procedure.

First, we define centrality classes in terms of the number of participants 
$N_{\rm p}$. Within a Glauber description (see Appendix~\ref{app:bdep1}), we define a centrality class by the threshold values $N^{\min}_{p}$ and $N^{\max}_{p}$ of $N_{\rm p}$ which saturate approximately the same fraction of the total interaction probability $P(N_{p} \ge 1) =1$ as the fraction of the total inelastic cross section attributed to the centrality classes in the experimental selection procedure. We note however that centrality in the experiment is defined in terms of event multiplicity rather than $N_{\rm p}$. In a given multiplicity class one may have events where $N_{\rm p}$ is slightly above or below the thresholds defined according to a sharp cut on the fraction of the total interaction probability as a function of $N_{\rm p}$. In order to account for this possibility we widen the $N_{\rm p}$ interval attributed to each class and use the new threshold values in further estimates. Those values are given in Table~\ref{tab:centr-Npart}.
As a consistency check of our class selection method, we calculate the average number of binary collisions $\ave{N_c}$ for the different centrality classes. These numbers coincide with the results of the Glauber Monte-Carlo supplemented with the \pp\ RHIC data used by the PHENIX collaboration (see Table 1 in Ref.~\cite{Adare:2012qf}). 

\begin{table}[t] 
\small
\centering
\renewcommand{\tabcolsep}{3pt}
\begin{tabular}{c|c|c|c|c||c|c|c|c|c}
\multicolumn{5}{c||}{Glauber, RHIC} & \multicolumn{5}{c}{Glauber, LHC}\\
\hline
class & $N^{\min}_{p}$; $N^{\max}_{p}$ & $\frac{P({\rm class})}{P(N\ge 1)}$ & $\ave{N_{c}}$ & $L_{\rm Au}$ & class & $N^{\min}_{p}$; $N^{\max}_{p}$ & $\frac{P({\rm class})}{P(N\ge 1)}$ & $\ave{N_{c}}$ & $L_{\rm Pb}$ \\
\hline
A &11; 197 & 0.28 & 15.9 & 12.87  & 1 & 12; 208 & 0.246 & 14.8 & 13.46 \\
B &8; 12 &  0.24 & 10.9 & 9.62    & 2 & 9; 12  & 0.215 & 10.5 & 9.55  \\
C &5; 8 & 0.23 & 7.0 & 7.17       & 3 & 5; 8 & 0.215 & 6.5 & 6.29 \\
D &2; 4 & 0.29 & 3.6 & 3.84       & 4 & 1; 5 & 0.428 & 2.4 & 3.39 \\
\end{tabular}

\caption{Average number of binary collisions $\ave{N_{c}}$, and average path length $L_{\rm A}$ (in fm) for different centrality classes (A--D at RHIC and 1--4 at LHC), as obtained in a Glauber calculation. See text and Appendix~\ref{app:bdep} for details.}
\label{tab:centr-Npart}
\end{table}

Second, within each of the centrality classes defined in such a way, we determine, also in the Glauber model, the average number of target nucleons participating in the rescattering of the fast color octet $Q \bar Q$ pair,
see Appendix~\ref{app:bdep2}. The average path length $L_{\rm A}$ for a given centrality class directly follows from this number. Table~\ref{tab:centr-Npart} displays the values of $L_{\rm Au}$ (RHIC) and $L_{\rm Pb}$ (LHC) to be used in our model, for each centrality class.   

\section{Comparison to  E866 and RHIC data}
\label{sec:comparison}

The $\jpsi$ nuclear production ratio is computed as a function of $\pt$ for various values of $y$ or $\xf = (2 M_\perp/\sqrt{s}) \, \sinh{y}$. 
The only parameter of the model, the transport coefficient $\qzero$, has been fixed to $\qzero=0.075$~\gevsqfm\ in~\cite{Arleo:2012rs} from the $\xf$ dependence of $\jpsi$ suppression measured in p--W collisions by E866~\cite{Leitch:1999ea}. In the numerical calculations we use $\lqcd = 0.25 \,{\rm GeV}$ (however see footnote~\ref{foot:lqcddep}), $\alpha_s =0.5$, and $M=  3 \,{\rm GeV}$ ($M=  9 \,{\rm GeV}$) for the mass of the $c\bar{c}$ ($b\bar{b}$) 
pair. 

\subsection{E866} 
\label{sec:comparisone866}

The E866 collaboration has measured the $\jpsi$ suppression in p--Fe and p--W collisions (with respect to p--Be) at $\sqrt{s}=38.7$~GeV as a function of the transverse momentum for three domains in $\xf$~\cite{Leitch:1999ea}. At small $\xf$, $\jpsi$ production can be affected by nuclear absorption since the typical $\jpsi$ hadronization 
time becomes comparable to (or less than) the size of the nuclear targets~\cite{Arleo:2012rs}. The model is 
therefore compared to the E866 data in the intermediate-$\xf$ ($0.2\leq\xf\leq0.6$, $\langle\xf\rangle=0.308$) and large-$\xf$ ($0.3\leq\xf\leq0.93$, $\langle\xf\rangle=0.48$) domains.

\begin{figure}[htbp]
\begin{center}
\includegraphics[width=12cm]{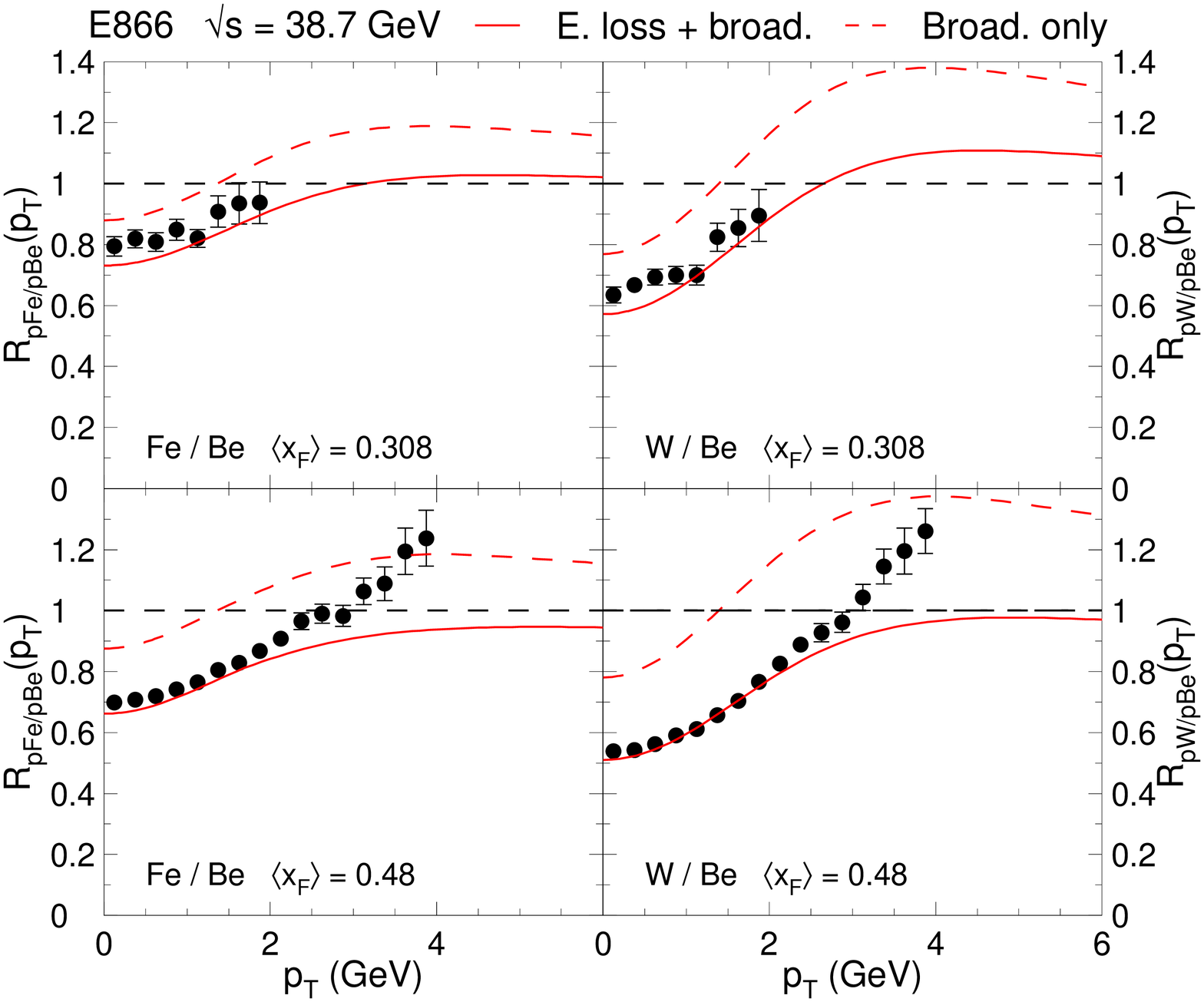}
\end{center}
\caption{Model predictions (solid red curves) for the $\jpsi$ nuclear suppression factor compared to the E866 data for $R_{\rm Fe/Be}(\pt)$ (left) and $R_{\rm W/Be}(\pt)$ (right), in the intermediate-$\xf$ ($\ave{\xf} = 0.308$, top) and large-$\xf$ ($\ave{\xf} = 0.48$, bottom) ranges. The dashed lines indicate the effect of momentum broadening only, $R_{\rm pA}^{\rm broad}(y, \pt)$, Eq.~\eq{eq:rpabroad}.}
\label{fig:E866}
\end{figure}

In Fig.~\ref{fig:E866} are shown as solid lines the Fe/Be (left) and W/Be (right) nuclear production ratios at intermediate (top) and large (bottom) $\xf$. The ratio $\RpA(\pt)$ increases with $\pt$ in almost the whole $\pt$ range, with more pronounced suppression (at $\pt=0$~GeV, where $\RpA$ is the smallest) in W targets and at large $\xf$. The $\pt$ dependence  essentially arises from that of $\RpA^{\rm broad}(\pt)$, shown as dashed lines.
The sole energy loss effect, $\RpA^{\rm loss}(y, \pt)$, proves rather flat in this $\pt$-domain, but is essential to fix the magnitude of $\RpA$, leading to 
a remarkable agreement between the data and the model predictions 
in the $0\leq\pt\leq2$~GeV range. 

At large $\xf$, the data overshoot the theoretical expectations above $\pt\gtrsim3$~GeV, for which the model predictions flatten out. Several reasons might explain this disagreement. First of all, the parametrization of the p--p cross section (using E789 data, see Section~\ref{sec:ppfits}) has been performed at $\xf=0$ only; it could therefore well be that this fit, used to compute $\RpA$, is no longer appropriate to describe the p--p production cross section at both large $\xf \simeq 0.4$ \emph{and} $\pt\gtrsim 3$~GeV. Also note that the theoretical calculations have been performed at $\xf=\langle\xf\rangle$ while the data are averaged on a rather large $\xf$-bin. When $\pt$ gets larger, the kinematical correlation between $\pt$ and $\xf$ would therefore tend to decrease the typical $\xf$, for which the model would predict slightly larger $\RpA$ ratios.\footnote{We checked that averaging properly over the whole $\xf$-range considered experimentally increases by $\sim 5\%$ the $\RpA$ ratios at large $\pt$, therefore slightly reducing the discrepancy with experimental data.} For those reasons it is difficult to draw any firm conclusion from the comparison between the model and the E866 data at large $\xf$ and $\pt$. 

Putting aside the latter $(\pt, \xf)$ region, it is remarkable that the model reproduces quantitatively the $\pt$ dependence of $\jpsi$ suppression in various nuclei and at different $\xf$ values. The fact that the {\it same} quantity, $\hat{q} L$, determines the strength of medium-induced gluon radiation (and therefore energy loss) necessary to explain the $\xf$ dependence of $\RpA$ (see Ref.~\cite{Arleo:2012rs}) and the amount of momentum broadening required in the present study to reproduce the shape  of $\RpA$ as a function of $\pt$,  
strongly supports parton energy loss induced by momentum broadening as the dominant effect in quarkonium nuclear suppression. 
 
\subsection{RHIC} 

Let us now move to RHIC energy, where the $\pt$ dependence of $\jpsi$ suppression in d--Au collisions has been reported recently by the PHENIX collaboration~\cite{Adare:2012qf}.

\begin{figure}[htbp]
\begin{center}
\includegraphics[width=4.8cm]{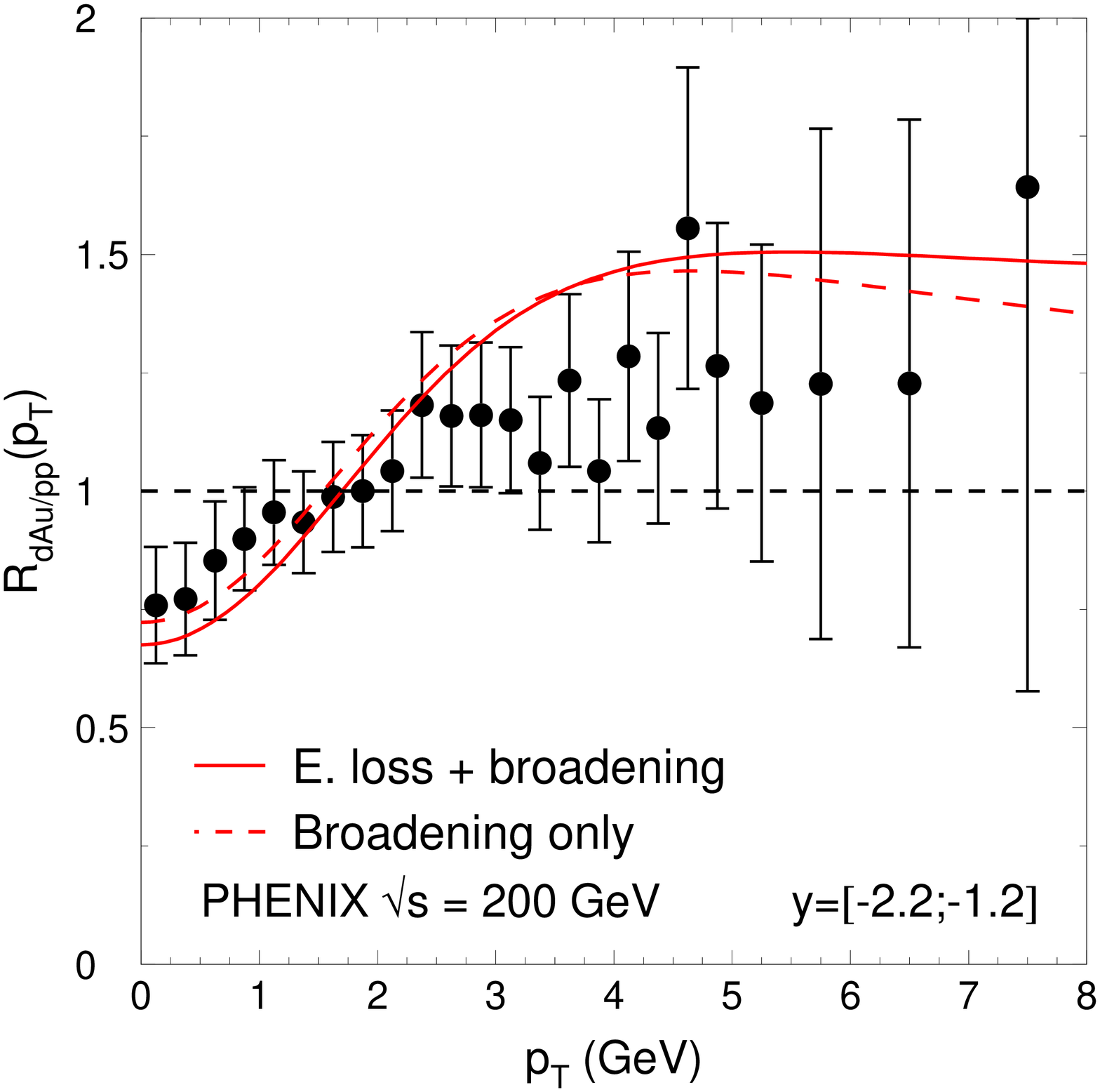}
\includegraphics[width=4.8cm]{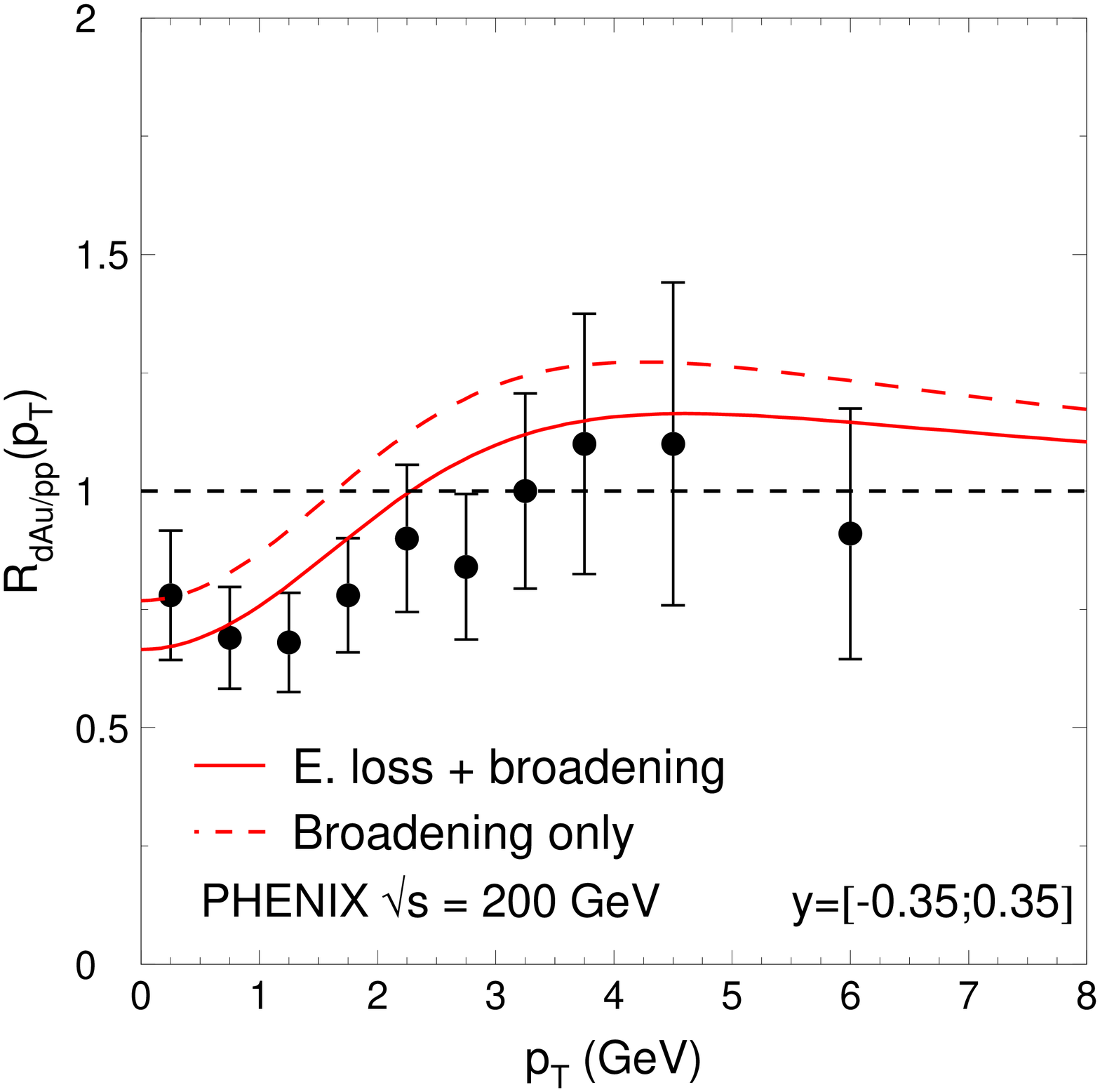}
\includegraphics[width=4.8cm]{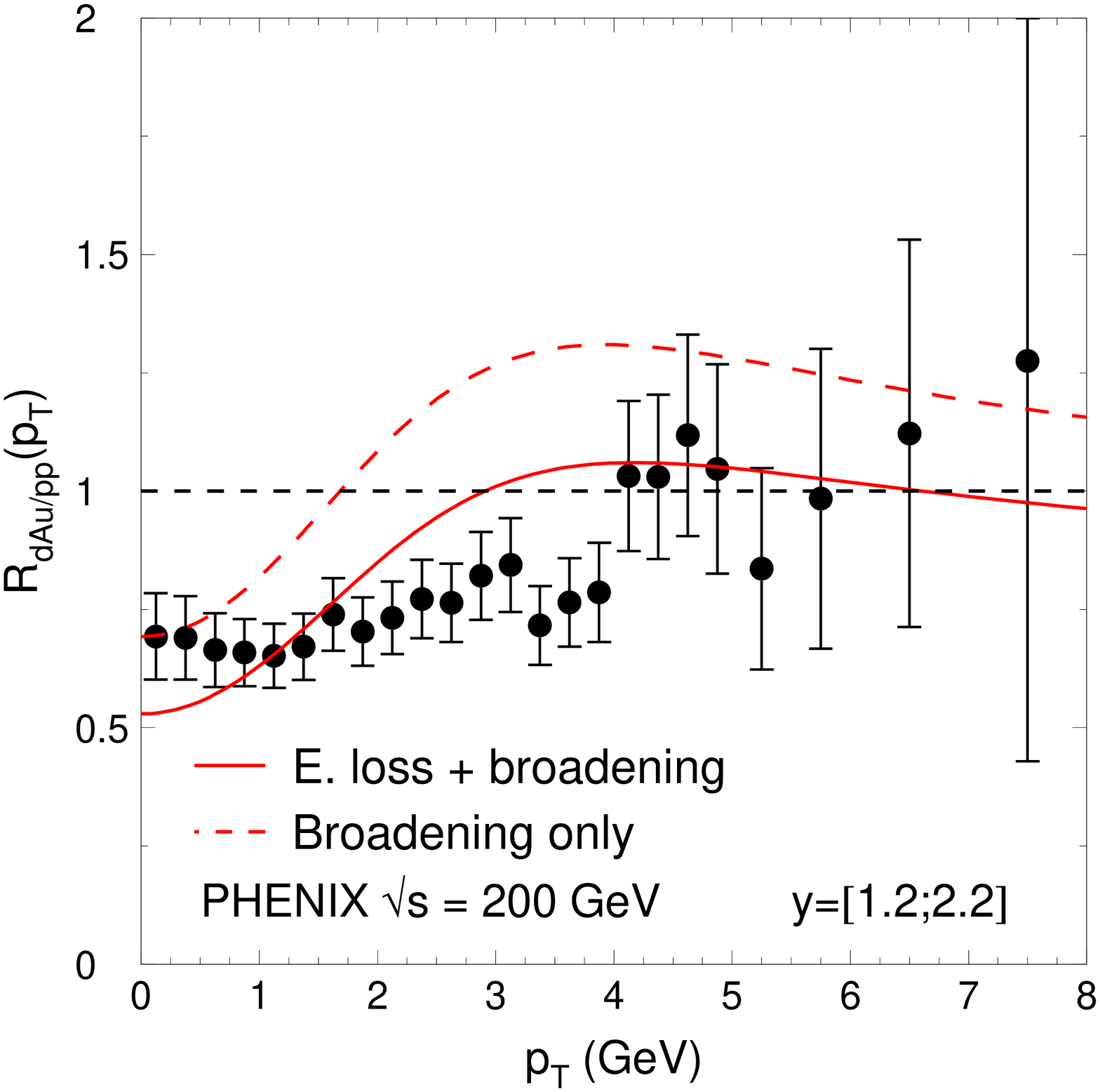}
\end{center}
\caption{Model predictions for the $\jpsi$ nuclear suppression factor $R_{\rm pA}(\pt)$ in minimum bias d--Au collisions at RHIC, at backward (left), central (middle) and forward (right) rapidities (solid curves). The dashed lines indicate the effect of momentum broadening only, $R_{\rm pA}^{\rm broad}(y, \pt)$, Eq.~\eq{eq:rpabroad}.}
\label{fig:RHIC-minbias}
\end{figure}

In Fig.~\ref{fig:RHIC-minbias} the model predictions are compared to the PHENIX data measured in minimum bias d--Au collisions, at backward ($-2.2\leq{y}\leq-1.2$, left), central ($|y|\leq0.35$, middle) and forward ($1.2\leq{y}\leq2.2$, right) rapidities.\footnote{Calculations are performed at the fixed values of $y=-1.7$, $0$, $+1.7$ respectively. We checked that similar results are obtained when averaging over the experimental $y$ range. Note also that for $y=-1.7$, the $\jpsi$ hadronization time is comparable to the size of the gold nucleus, and nuclear absorption may play a role~\cite{Arleo:2012rs}.} The model reproduces the trend seen in data, namely an increase with $\pt$ up to $\pt\simeq4$--$5$~GeV. Around those values, some nuclear enhancement, $\RpA(\pt)>1$, is visible at backward and central rapidities, but would need more precise data to be confirmed. At forward rapidity, the suppression due to energy loss is too strong to observe such an enhancement, both in the data and in the model.
As for the results at E866 energy discussed earlier, the $\pt$ shape  
of $\RpA(\pt)$ is essentially driven by the effect of momentum broadening, Eq.~\eq{eq:rpabroad}, shown as dashed lines.
   
\begin{figure}[htbp]
\begin{center}
\includegraphics[width=7.5cm]{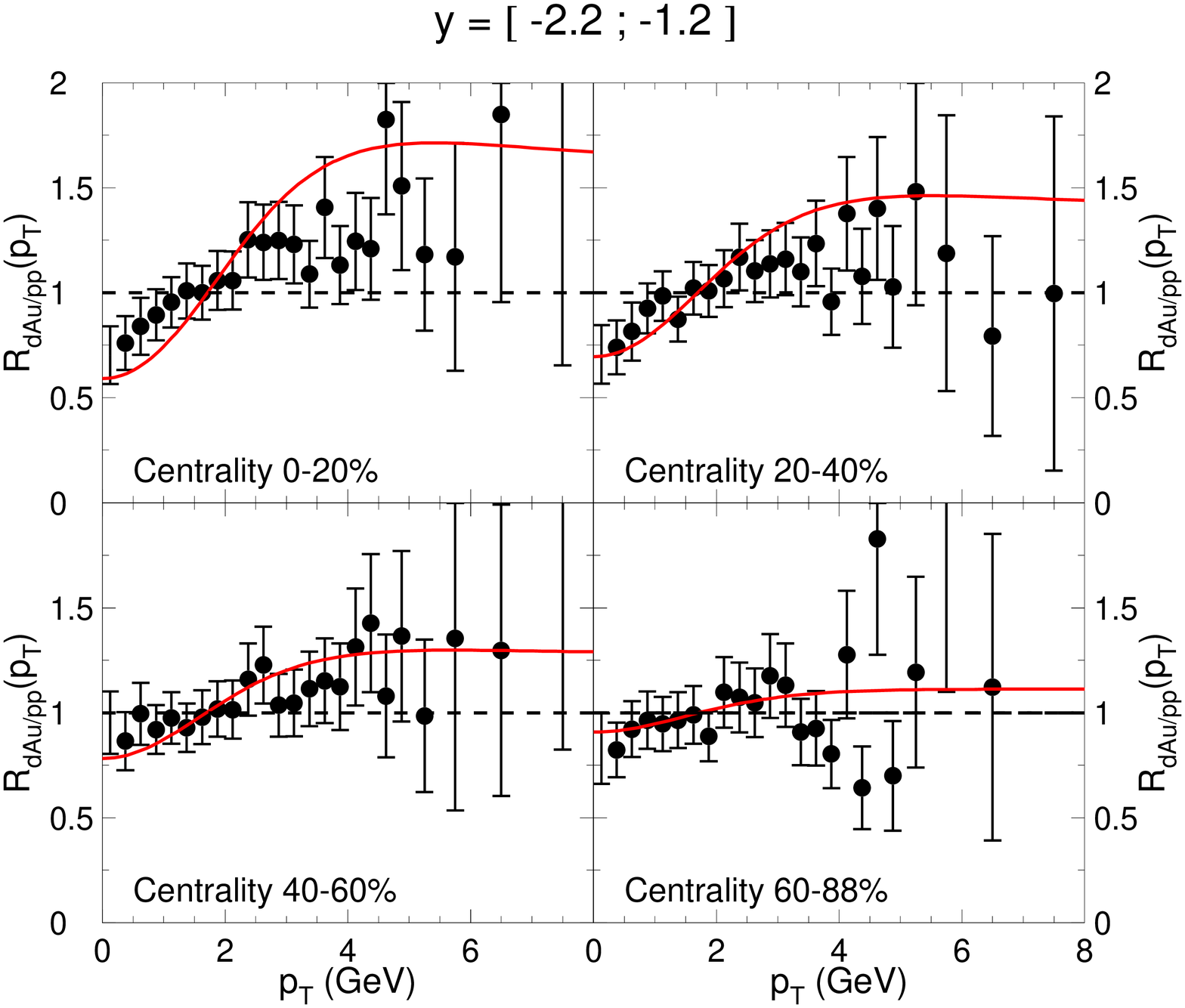}
\includegraphics[width=7.5cm]{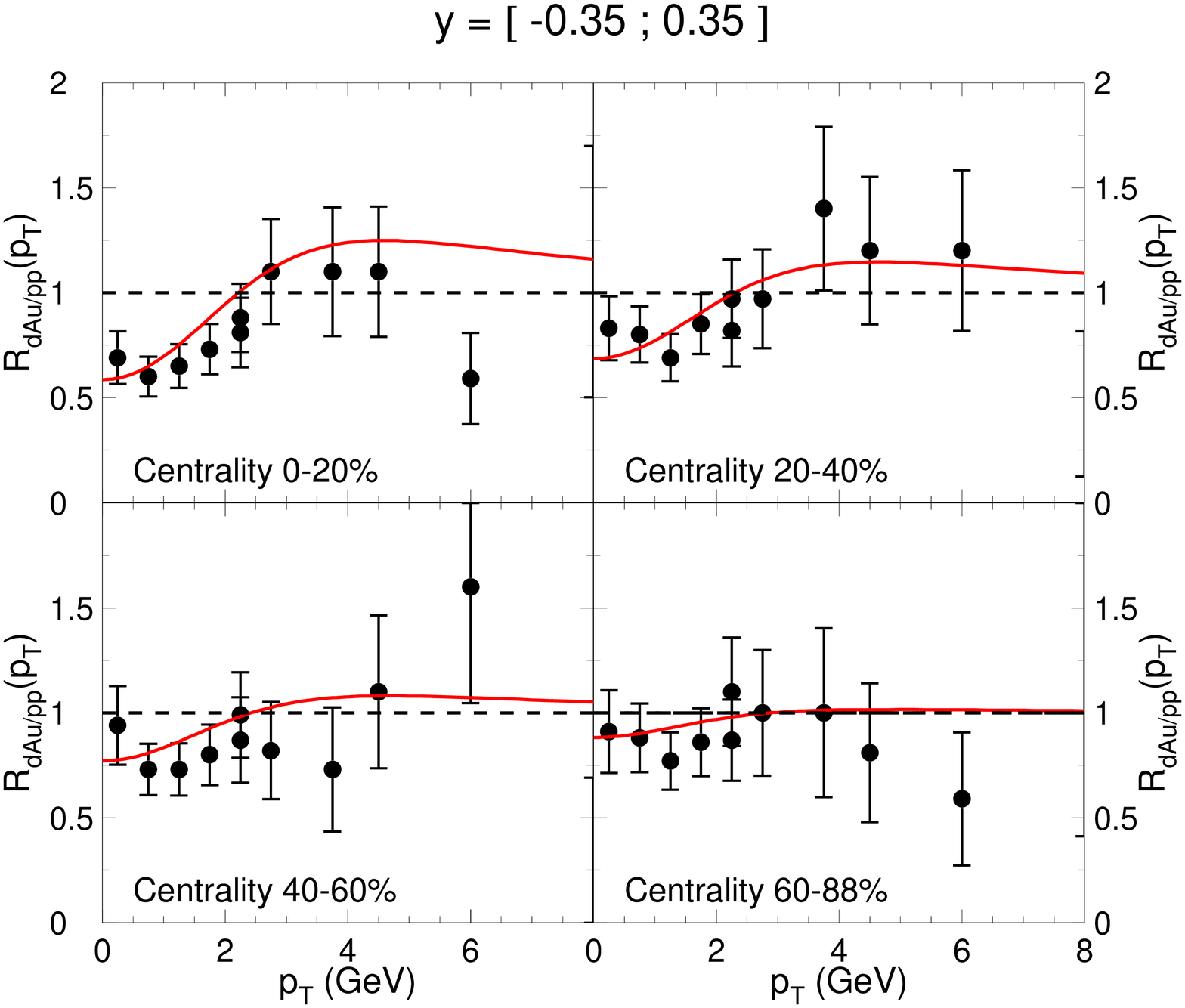}
\includegraphics[width=7.5cm]{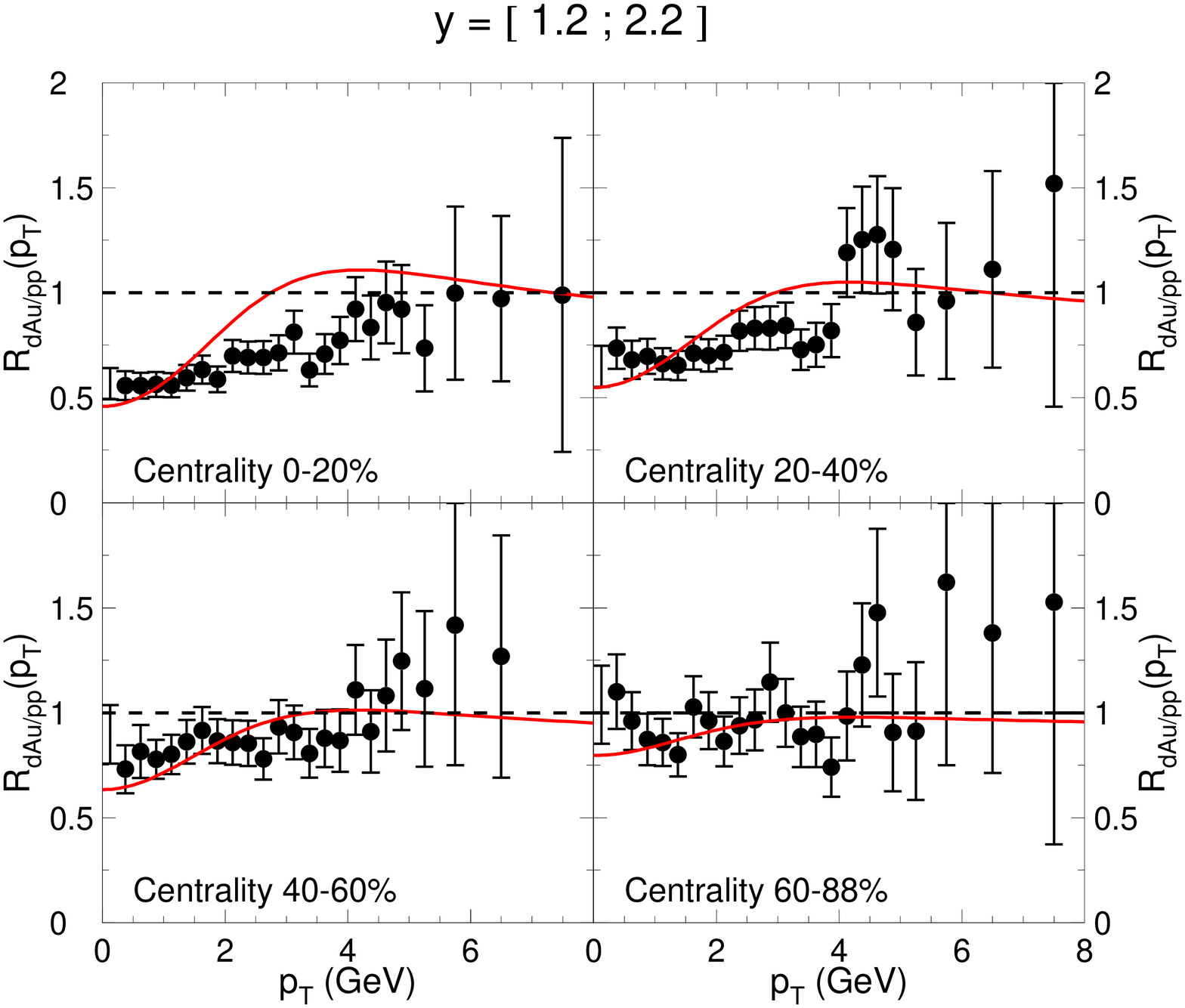}
\end{center}
\caption{Same as Fig.~\ref{fig:RHIC-minbias} in the four centrality classes (from top left to bottom right in each panel) at backward, central and forward rapidities (upper left, upper right, bottom).}
\label{fig:RHIC-centrality}
\end{figure}

On top of minimum bias collisions, $\jpsi$ suppression has also been measured in four centrality classes of d--Au collisions ($0$--$20\%$, $20$--$40\%$, $40$--$60\%$, $60$--$88\%$). The data are shown in Fig.~\ref{fig:RHIC-centrality} for the three rapidity bins and four centrality classes and compared to the model. 
Details on the medium length $L$ corresponding to the various centrality classes of d--Au collisions and used in the theoretical calculation can be found in Section~\ref{sec:bdep} and Appendix~\ref{app:bdep}. As for minimum bias collisions, the model predictions prove in very good 
agreement with data. In particular, the centrality dependence is well reproduced by the model, with rather pronounced effects in the $0$--$20\%$ most central collisions, and almost negligible effects, $\RpA\simeq1$ in the whole $\pt$-range for the $60$--$88\%$ most peripheral collisions.

Therefore it appears that the present model, based on parton energy loss and momentum broadening, offers a better agreement on the $\pt$ and centrality dependence of $\RpA$ than models based purely on modification of parton densities~\cite{Ferreiro:2012sy} and nuclear absorption~\cite{Kopeliovich:2011zz} which tend to predict a flatter dependence of $\RpA(\pt)$.

\section{Predictions for p--Pb collisions at the LHC} 
\label{sec:LHCpredictions}

A run of p--Pb collisions at $\sqrts=5$~TeV has taken place at LHC early 2013 with an integrated luminosity of ${\cal L}\simeq30$~nb$^{-1}$ collected by ALICE, ATLAS and CMS experiments. 
This will allow for precise measurements of $\jpsi$ and $\Upsilon$ production in p--A collisions at an unprecedented energy, on a wide range of rapidities ($|y|<5$) and transverse momenta, and consequently clarify the role of cold nuclear matter effects at high energy.
In this section, we therefore provide predictions for $\jpsi$ and $\Upsilon$  nuclear production ratios $\RpA$ as a function of $\pt$, for different 
rapidities ($y=-3.7$, $0$, $2.8$)\footnote{Here the rapidity $y$ is defined in the center-of-mass frame of the proton--nucleon collision, related to the rapidity in the laboratory frame $y_{\rm lab}$ as $y = y_{\rm lab} - 0.465$ in p--Pb collisions and $y = y_{\rm lab} + 0.465$ in Pb--p collisions. The value $y=2.8$ (respectively $y=-3.7$) is chosen to correspond to the median rapidity of the ALICE spectrometer acceptance, namely $2.5<y_{\rm lab}<4$, in p--Pb (respectively Pb--p) collisions.}  and centrality classes (labelled 1\dots4 from central to peripheral collisions, see Section~\ref{sec:bdep} for details).

\begin{figure}[ht]
\begin{center}
\includegraphics[width=4.8cm]{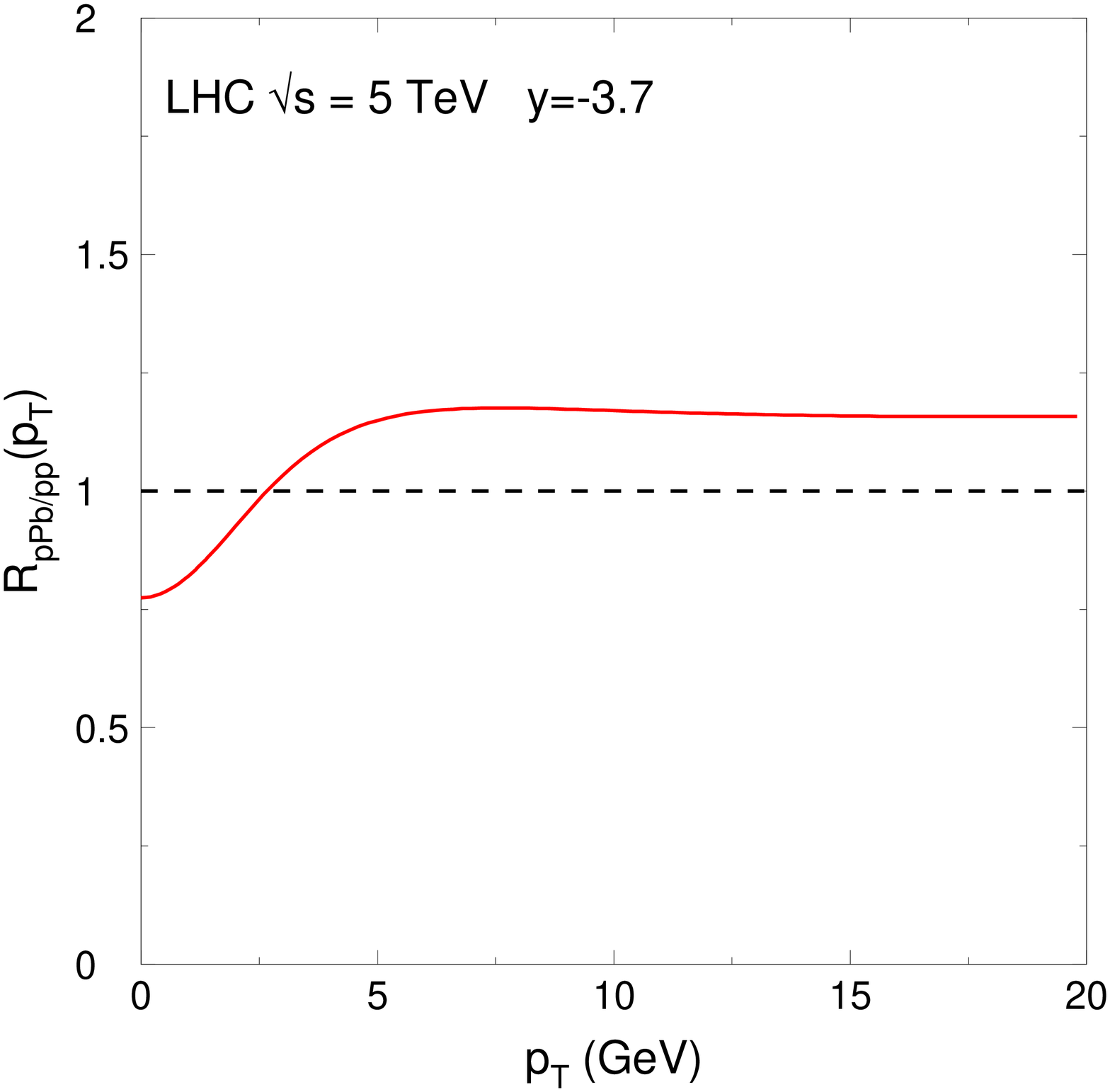}
\includegraphics[width=4.8cm]{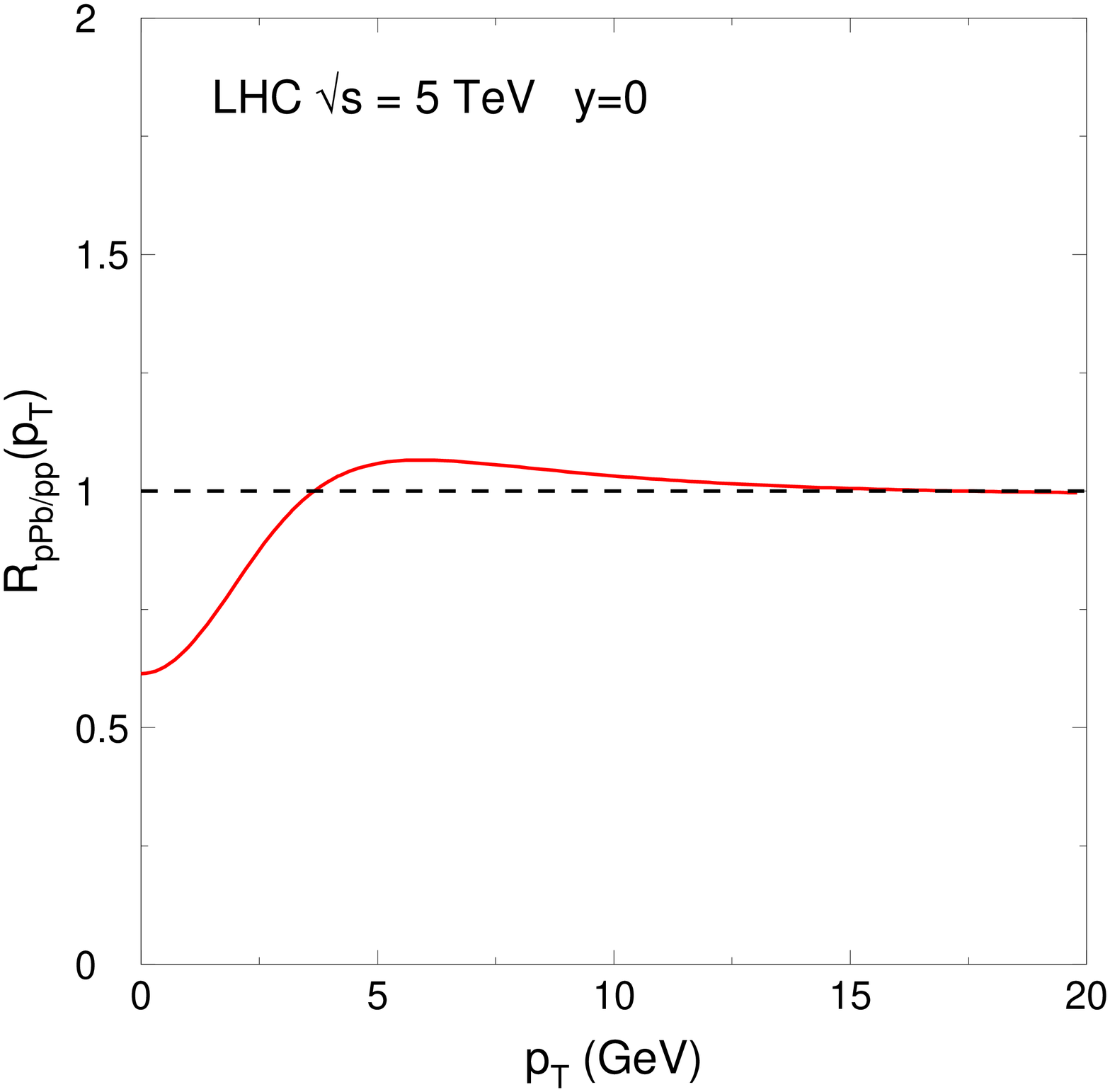}
\includegraphics[width=4.8cm]{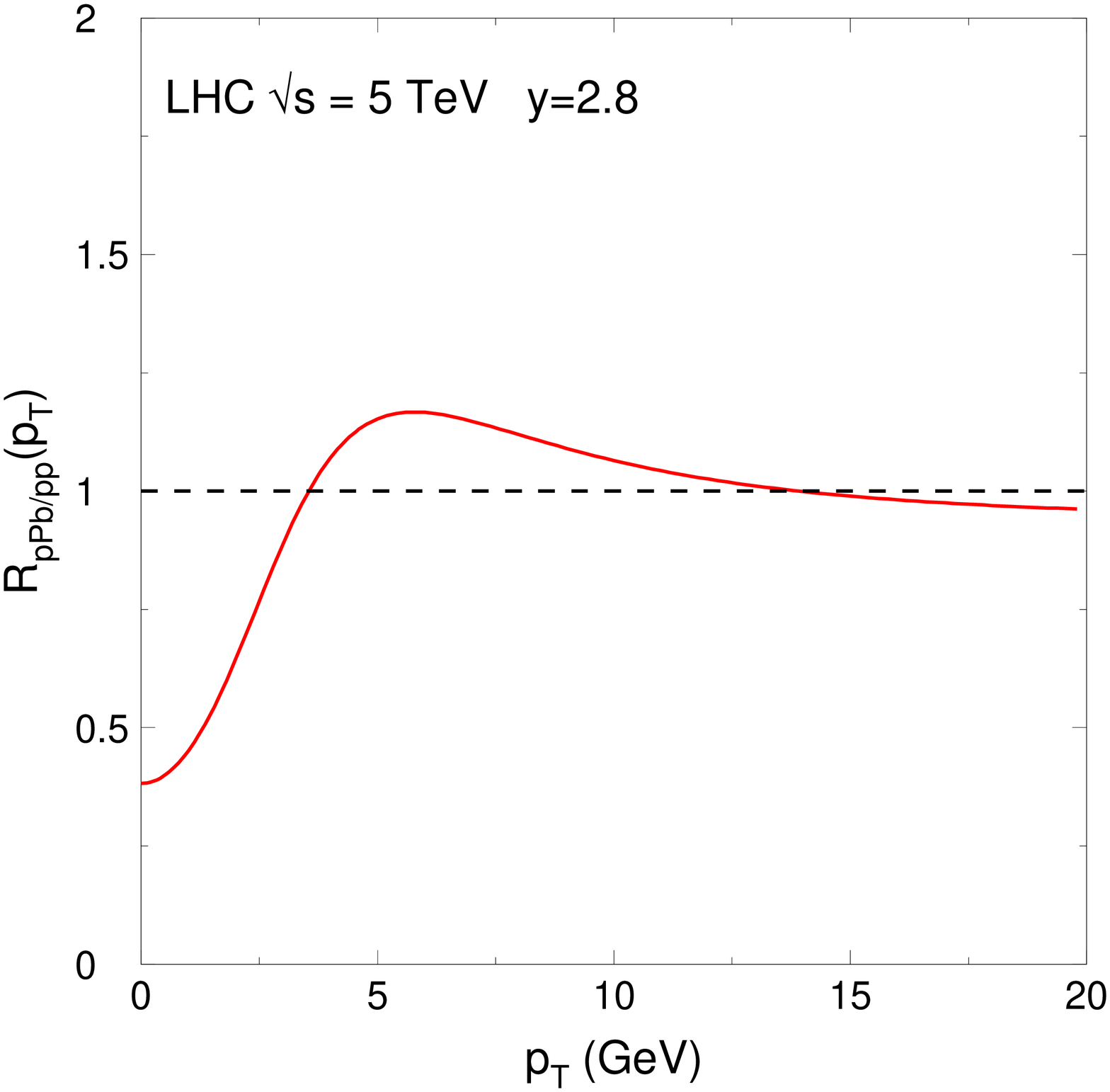}
 \end{center}
\caption{Prediction for the $\jpsi$ nuclear suppression factor $R_{\rm pA}(\pt)$ in minimum bias p--Pb collisions at LHC, for backward, central and forward rapidities.}
\label{fig:LHC}
\end{figure}

In Fig.~\ref{fig:LHC} we show the $\jpsi$ nuclear production ratio for minimum bias p--Pb collisions, from backward (left) to central (middle) and forward (right) rapidities. At all rapidities a depletion of $\jpsi$ production is expected at low $\pt$, say $\pt \lesssim 3$~GeV. At larger $\pt$ a ``Cronin peak'' might only be visible at large rapidity, $y=2.8$ and above, in minimum bias collisions. Quite generally the effects of both energy loss and momentum broadening are expected to become more pronounced at larger rapidities. As shown in~\cite{Arleo:2012rs}, energy loss effects prove stronger at large positive rapidity because of the energy dependence of the average energy loss, $\Delta E \propto E$, associated to the medium-induced spectrum \eq{spectrum}. 

\begin{figure}[t]
\begin{center}
\includegraphics[width=7.5cm]{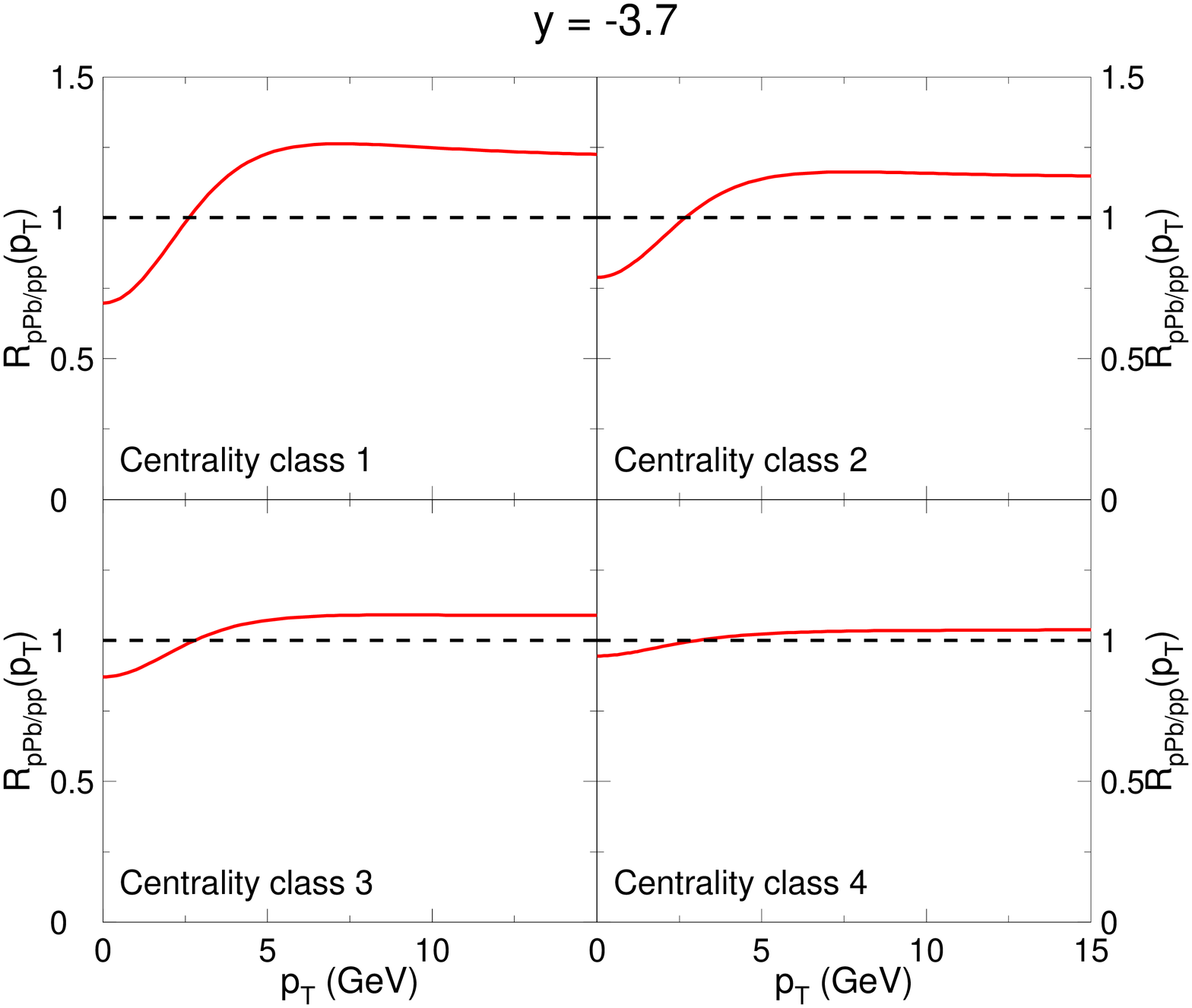}
\includegraphics[width=7.5cm]{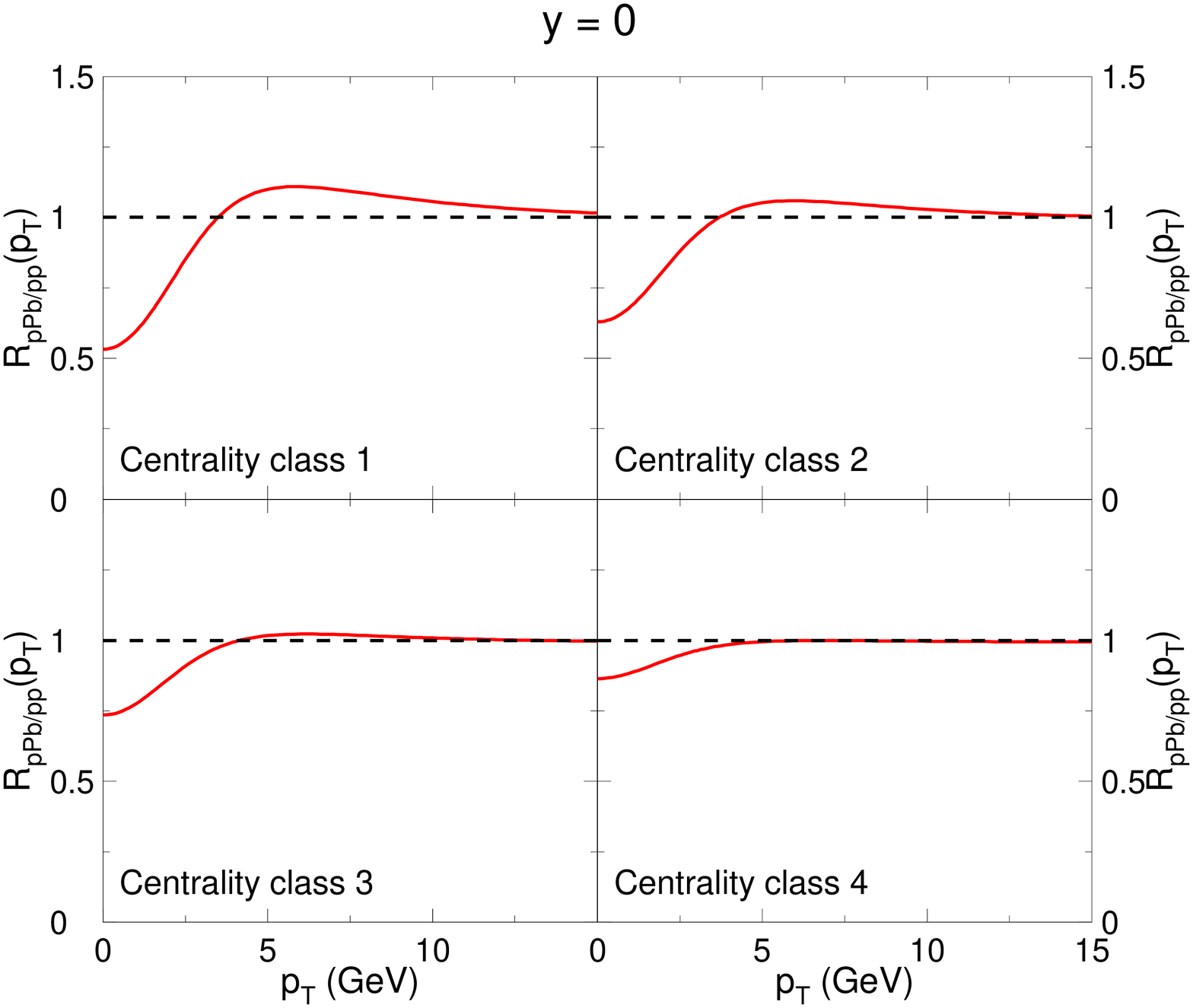}
\includegraphics[width=7.5cm]{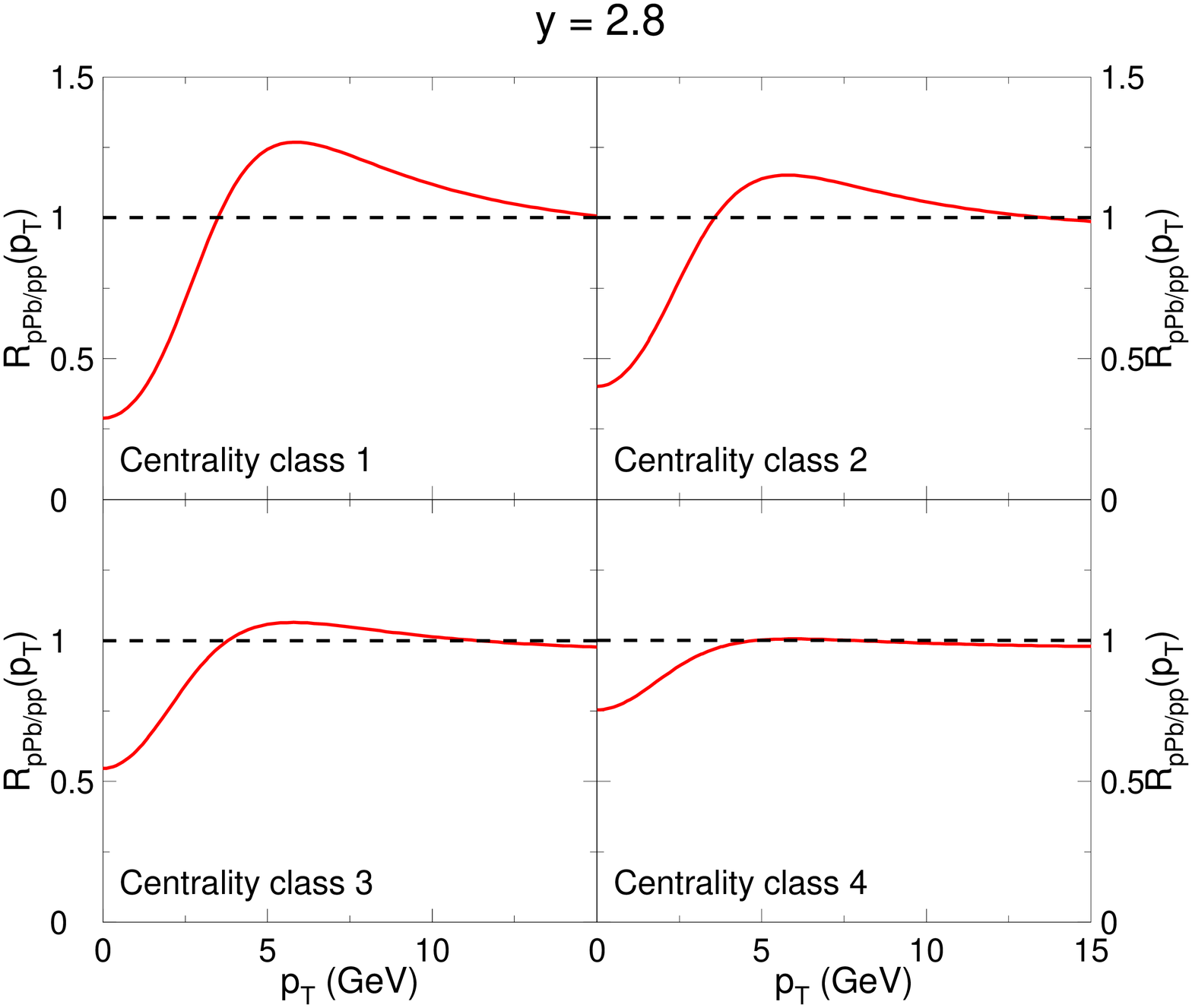}
\end{center}
\caption{Same as Fig.~\ref{fig:LHC} in the four centrality classes (from top left to bottom right in each panel) at backward, central and forward rapidities (upper left, upper right, bottom).}
\label{fig:LHC-centrality}
\end{figure}

The model predictions are also provided in the four centrality classes of p--Pb collisions in Fig.~\ref{fig:LHC-centrality}. As expected the deviations of $\RpA$ from unity are largest in the most central collisions, while in the most peripheral p--Pb collisions (centrality class 4), $\RpA(\pt)\simeq1$ at all $\pt\gtrsim2$--$3$~GeV. 
The most spectacular effects can be seen in the central collisions (class 1) and at forward rapidity, where 
$\RpA\simeq0.25$ at $\pt=0$~GeV and $\RpA\simeq 1.3$ at $\pt=6$~GeV.

\begin{figure}[h]
\begin{center}
\includegraphics[width=6cm]{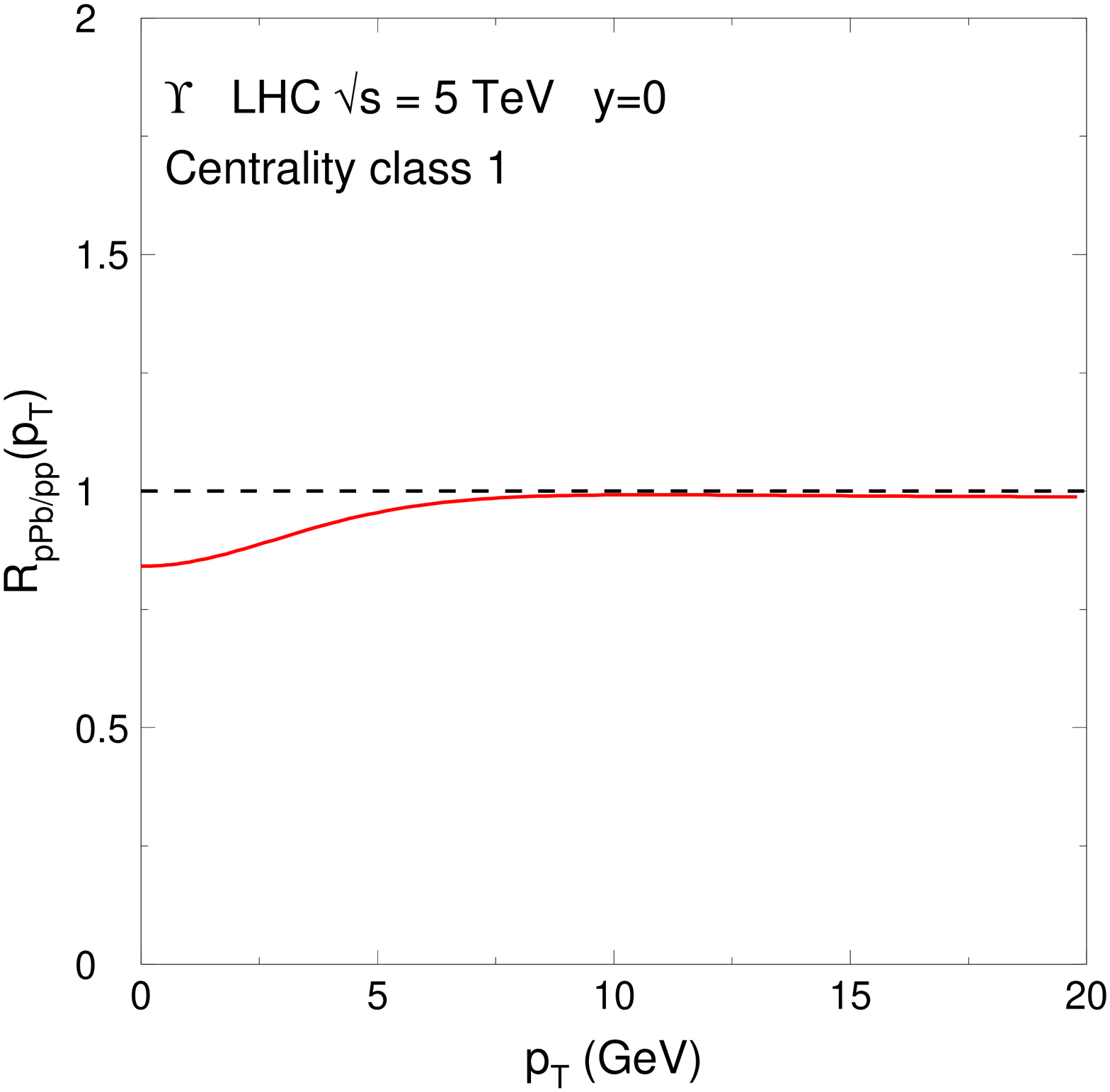}
\includegraphics[width=6cm]{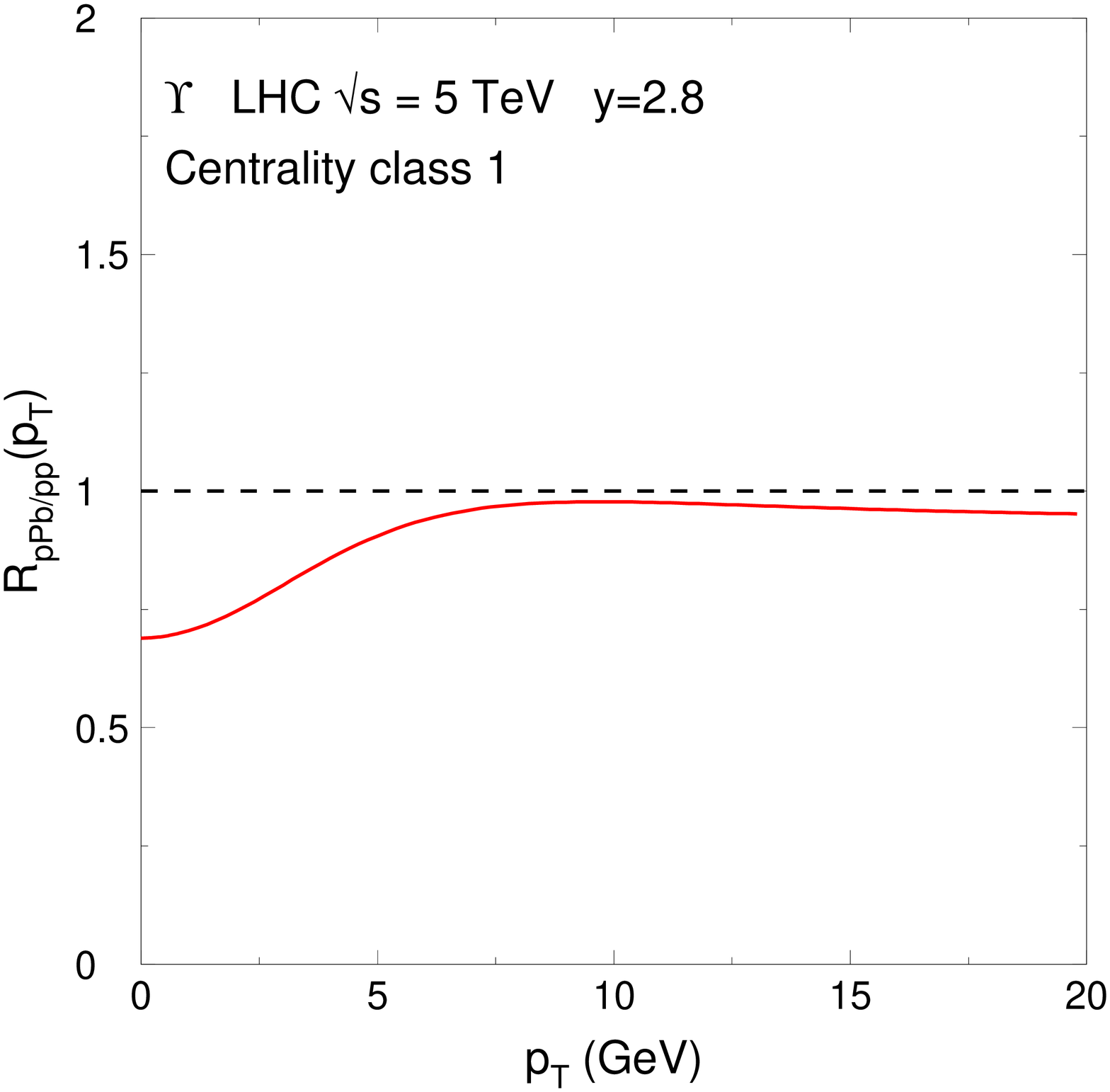}
\end{center}
\caption{Predictions for the $\Upsilon$ nuclear suppression factor $R_{\rm pA}(\pt)$ in central p--Pb collisions (centrality class 1) at LHC, for central (left) and forward (right) rapidities.}
\label{fig:LHC-ups}
\end{figure}

Predictions are performed as well in the $\Upsilon$ channel. The expected $\RpA^{\Upsilon}$ at the LHC is shown in Fig.~\ref{fig:LHC-ups} for the most central p--Pb collisions (centrality class 1). Because of the mass dependence of the energy loss, 
$\Delta E \propto \mT^{-1}$, the suppression due to energy loss, $\RpA^{\rm loss}$, is milder than for $\jpsi$. Moreover, although the amount of momentum broadening experienced by $\jpsi$ and $\Upsilon$ states is expected to be similar (neglecting the $x$ dependence of the transport coefficient), the height of the Cronin peak is much less pronounced for $\Upsilon$ than for $\jpsi$ (compare \eg\ Fig.~\ref{fig:LHC-centrality} bottom and Fig.~\ref{fig:LHC-ups} right).\footnote{As a matter of fact, the slight enhancement arising from $\RpA^{\rm broad}$ is compensated by energy loss effects, $\RpA^{\rm loss}<1$, making $\RpA^\Upsilon$ smaller than one at all $\pt$.} 
This is due to 
the flatter $\Upsilon$ $\pt$-spectrum as compared to that of $\jpsi$ production, see Fig.~\ref{fig:fit2d} and the values of the parameters $p_0$ and $m$ in Table~\ref{tab:param}. As can be seen in Fig.~\ref{fig:LHC-ups},  $\Upsilon$ suppression, $\RpA^{\Upsilon}<1$, is predicted in the range $0\leq\pt\leq6$~GeV at mid-rapidity. The suppression extends to larger $\pt$ than for $\jpsi$ due to the larger value of the $p_0$ parameter in the p--p cross section. The suppression is maximal at $\pt=0$~GeV, where $\RpA^\Upsilon\simeq0.85$ (resp. $0.65$) at $y=0$ (resp. $y=2.8$).

\section{Conclusion}\label{sec:conclusion}

Following our earlier work~\cite{Arleo:2012rs}, we studied the effects of parton $\pt$-broadening and energy loss in cold nuclear matter on the $\pt$ dependence of $\jpsi$ and $\Upsilon$ suppression in p--A collisions. We found that the momentum broadening is responsible for the fast variation of $\jpsi$ suppression with $\pt$, while medium-induced energy loss essentially affects the magnitude of $\RpA$.

Using the transport coefficient $\qzero=0.075$~\gevsqfm\ fixed in~\cite{Arleo:2012rs}, the model predictions prove in very good agreement with recent PHENIX data~\cite{Adare:2012qf} in minimum bias and centrality-dependent d--Au collisions at $\sqrts=200$~GeV. Our results are also successfully compared to earlier results from the E866 collaboration~\cite{Leitch:1999ea}. 
Finally, predictions for $\jpsi$ and $\Upsilon$ suppression in p--Pb collisions (minimum bias and in four centrality classes) at the LHC ($\sqrts=5$~TeV) are provided.

The good description, within a consistent framework, of both the rapidity and transverse momentum dependence of $\RpA^{\jpsi}$ from fixed-target experiments to RHIC is a hint that 
parton energy loss induced by momentum broadening might 
be the dominant effect responsible for $\jpsi$ suppression in p--A collisions.

\acknowledgments
This work is funded by ``Agence Nationale de la Recherche'', 
grant ANR-PARTONPROP.  
R.~K.~acknowledges the Russian Foundation for Basic Research, project 12-02-00356-a.

\appendix

\section{Fits of \pp\  single and double differential rates}
\label{app:compare-param}

Here we compare the parametrization \eq{pp-fit-2d} for the double differential \pp\ cross section used in the present study and that for the single differential rate $\dd\sigma_{\mathrm{pp}}/\dd \xf$ used in~\cite{Arleo:2012rs},
\be
\label{pp-fit-xf}
\frac{\dd\sigma_{\mathrm{pp}}^{\psi}}{\dd\xf} \propto \frac{(1-x^\prime)^{n}}{x^\prime} \, , 
\ee
where the variables $\xf$ and $x^\prime$ are defined by
\bea
\label{xfdef}
\xf &=& \frac{E}{E_{\mathrm{p}}} - \frac{E_{\mathrm{p}}}{E} \, \frac{M_\perp^2}{s} = \frac{2 M_\perp}{\sqrt{s}} \, \sinh{y}  \, , \\
x^\prime  &=& \frac{E}{E_{\mathrm{p}}} + \frac{E_{\mathrm{p}}}{E} \, \frac{M_\perp^2}{s} = \frac{2 M_\perp}{\sqrt{s}} \, \cosh{y}  \, .
\eea
Eq.~\eq{pp-fit-xf} translates in rapidity to (use $\partial \xf / \partial y = x^\prime$)
\be
\label{pp-fit-rap}
\frac{\dd\sigma_{\mathrm{pp}}^{\psi}}{\dd y } \propto (1-x^\prime)^{n} = \left(1- \frac{2 M_\perp}{\sqrt{s}} \cosh{y} \right)^{n} = \nu(y, p_\perp) \, .
\ee
In Ref.~\cite{Arleo:2012rs} no information on $\pt$-distributions was used, and $\pt$ was replaced by some typical value, assumed to be $\bar{p}_\perp = 1 \,{\rm GeV}$. 

In the present study, the single differential cross section $\dd\sigma_{\mathrm{pp}}/\dd y$ can be obtained by integrating \eq{pp-fit-2d},
\be
\label{approx-single}
\frac{\dd\sigma_{\mathrm{pp}}^{\psi}}{\dd y } = \int_{\p} \frac{\dd\sigma_{\mathrm{pp}}^{\psi}}{\dd y\,\dd^2 \p} = {\cal N} \int_{\p} \mu(p_\perp) \, \nu(y, p_\perp) \simeq {\cal N} \left[ \int_{\p} \mu(p_\perp) \right] \,\nu(y, \bar{p}_\perp) \, , 
\ee
where the latter approximation arises from $\mu(p_\perp)$ decreasing much faster than $\nu(y, \pt)$ with $\pt$.
This can be checked for the values of the parameters $p_0$, $m$, $n$ (see Table~\ref{tab:param}) and within the intervals in $\pt$ and $y$ considered in our study (see Sections~\ref{sec:comparison} and~\ref{sec:LHCpredictions}). 
Comparing \eq{pp-fit-rap} and \eq{approx-single} we see that the parametrization \eq{pp-fit-2d} is consistent with that for the single differential rate \eq{pp-fit-rap} (or equivalently \eq{pp-fit-xf}) used in~\cite{Arleo:2012rs}. 

Let us remark that in \eq{approx-single} the typical $\bar{p}_\perp$ may be defined as
\be
\int_0^{\bar{p}_\perp^2} d\pt^2 \,\mu(p_\perp) \equiv \frac{1}{2} \int_0^{\infty} d\pt^2 \,\mu(p_\perp)  \Rightarrow \bar{p}_\perp^2 =  \left(2^{^{\frac{1}{m-1}}} -1 \right) p_0^2  \ \ \ (m>1) \, .
\label{medianpt}
\ee
Using the values of $p_0$ and $m$ given in Table~\ref{tab:param} for $\jpsi$, we find that $\bar{p}_\perp$ varies in the range $\bar{p}_\perp(p_0,m) = 1.3-2.4 \,{\rm GeV}$, somewhat above the {\it ad hoc} value $\bar{p}_\perp = 1 \,{\rm GeV}$ used in~\cite{Arleo:2012rs}. However, taking $\bar{p}_\perp \simeq 2 \,{\rm GeV}$ instead of $1 \,{\rm GeV}$ affects only slightly the value of $M_\perp$ ($M_\perp \simeq 3.6\,{\rm GeV}$ instead of $3.2\,{\rm GeV}$), with no sizeable effect on the predictions presented in Ref.~\cite{Arleo:2012rs}. 

\vspace{15mm} 

\section{Effective path length vs. centrality class}
\label{app:bdep}

In this Appendix we explain how the numbers of Table~\ref{tab:centr-Npart} in Section~\ref{sec:bdep} have been obtained.

\subsection{Number of participants and binary collisions for a given centrality class}
\label{app:bdep1}

\subsubsection*{\pA\ collisions (LHC case)}

In the Glauber model, at a given impact parameter $\b$ the number of participating nucleons $N_{\rm p}$ in the target nucleus follows a binomial distribution, 
\be
\label{eq:PNpartpA}
 P(N_{\rm p})=\cfrac{\int \dd {\b} \binom{A}{N_{\rm p}}  \left[ p_{\rm pA}(\b) \right]^{N_{\rm p}}  \left[ 1-p_{\rm pA}(\b) \right]^{A-N_{\rm p}}}{\int \dd {\b}\left\{1-   \left[ 1-p_{\rm pA}(\b) \right]^{A}\right\} } \, .
\ee
Here $p_{\rm pA}(\b) \equiv \sigma_{\rm in}^{\rm NN} \, T_{\rm A}(\b)$ is the probability for an inelastic collision between the projectile proton and a nucleon of the target nucleus, and $\sigma_{\rm in}^{\rm NN}$ is the inelastic \pp\  cross section which we take to be $70\,{\rm mb}$ at LHC energies. The target nucleus optical thickness $T_{\rm A}(\b)$ is normalized as $\int \dd \b \, T_{\rm A}(\b)=1$. The denominator in \eqref{eq:PNpartpA} is the p--A total inelastic cross section $\sigma_{\rm in}^{\rm pA}$ and ensures the correct normalization $P(N_{\rm p}>0)=1$ for the total interaction probability.

The number of binary collisions coincides with the number of participating nucleons of the target.

\subsubsection*{d--A collisions (RHIC case)}

The generalization of Eq.~\eqref{eq:PNpartpA} to the d--A case is straightforward, 
\be
\label{eq:PNpartdA}
P(N_{\rm p}) =  \cfrac{\int \dd \b \int \dd \r  P_{\rm d}({\r})   \binom{A}{N_{\rm p}}  \left[ p_{\rm dA}(\b,\r) \right]^{N_{\rm p}} \left[1- p_{\rm dA}(\b,\r) \right]^{A-N_{\rm p}}}{\int \dd \b \int \dd \r  P_{\rm d}(\r)  \left\{ 1 -  \left[1- p_{\rm dA}(\b,\r) \right]^{A}\right\}} \, ,
\ee
and is obtained by introducing the distribution $P_{\rm d}(\r)$ for the p--n transverse separation $\r$ in deuterium, and replacing $p_{\rm pA}(\b)$ by the interaction probability $p_{\rm dA}(\b,\r)$ of a target nucleon with either nucleon of the deuterium projectile, given by
\be\label{eq:pda}
p_{\rm dA}(\b,\r) \equiv \sigma_{\rm in}^{\rm NN} \,T_{\rm A}(\b+\tfrac{\r}{2}) + \sigma_{\rm in}^{\rm NN} \, T_{\rm A}(\b-\tfrac{\r}{2}) - p_{2}({\bf b},{\bf r}).
\ee
Here  
\be
\label{eq:p2}
p_{2}({\bf b},{\bf r}) \equiv \int \dd  {\bf s} T_{\rm A}({\bf s}) p({\bf b}+ {\bf r}/2-{\bf s}) p({\bf b}- {\bf r}/2 -{\bf s}) \simeq T_{\rm A}({\bf b})\int \dd {\bf s} p({\bf r}+{\bf s}) p({\bf s})
\ee
is the collision probability of a target nucleon with both nucleons of the deuterium projectile. The denominator of \eqref{eq:PNpartdA} is the d--A total inelastic cross section $\sigma_{\rm in}^{\rm dA}$.

The probability \eqref{eq:p2} depends on the N--N inelastic collision probability as a function of the impact parameter $p({\bf s})$. For that profile at c.m. energy $\sqrt s = 200$~GeV we take the Regge-inspired parametrization $p(\b)=1-\exp(-2Ne^{-{\bf b}^2/\alpha})$ with $\alpha = 1.05$~fm$^{2}$ and $N=1.1$, giving the total \pp\  inelastic cross section $\sigma_{\rm in}^{\rm NN}=\int \dd \s \, p(\s) = 4.2 \, {\rm fm}^2 = 42\,{\rm mb}$. The distribution $P_{\rm d}(\r)$ is evaluated by assuming a Hulthen form for the deuterium wave function. The thickness functions of all target nuclei considered in our study (Be, Fe, W, Au, Pb) are extracted from low-energy electron--proton scattering experiments~\cite{DeJager:1987qc}.

Summing up the probabilities \eq{eq:PNpartpA} and \eq{eq:PNpartdA} we find the threshold values $N^{\min}_{p}$ and $N^{\max}_{p}$ 
which saturate approximately 20\% of the total interaction probability $P(N_{\rm p}\ge 1) =1$, as described in Section~\ref{sec:bdep}.

In the d--A case, the number of binary collisions differs from $N_{\rm p}$ by the number of target nucleons which undergo collisions with both nucleons of the deuterium projectile:
\begin{equation}
\label{eq:dNcoll}
 \langle \Delta N \rangle|_{N\in[N_1,N_2]} = \cfrac{\sum\limits_{N=N_1}^{N_2}N \binom{A}{N}  \int \dd{\bf b}   \int \dd{\bf r}   P_{\rm d}({\bf r})     p_{2}({\bf b},{\bf r})\left[ p_{\rm dA}({\bf b},{\bf r}) \right]^{N-1} \left[1- p_{\rm dA}({\bf b},{\bf r}) \right]^{A-N}}
 {\sum\limits_{N=N_1}^{N_2}\binom{A}{N}  \int \dd{\bf b}   \int \dd{\bf r}   P_{\rm d}({\bf r})   \left[ p_{\rm dA}({\bf b},{\bf r}) \right]^{N} \left[1- p_{\rm dA}({\bf b},{\bf r}) \right]^{A-N}}.
\end{equation}
\subsection{Average number of collisions in the events with a hard process}
\label{app:bdep2}

\subsubsection*{\pA\ (LHC case)}

In the case of triggering on $\jpsi$ production (or any other hard process with a small cross section) the average number of collisions for the participating nucleon gets modified. The hard production process can occur in each of the inelastic collisions of the projectile nucleon with a probability $\sigma_{\jpsi}/\sigma_{\rm in}^{\rm NN}$,\footnote{We assume that the hard process cross section is small, so that the probability of two hard processes in the same \pA\ collision can be neglected.} so that the hard process probability is $P(\jpsi|N_{\rm p}) = N_{\rm p} \, \sigma_{\jpsi}/\sigma_{\rm in}^{\rm NN}$ and the joint probability for the hard production and $N_{\rm p}$ participating nucleons in the target is
\be
\label{eq:PhardNp}
 P(N_{\rm p},\jpsi)= \frac{N_{\rm p}}{\sigma_{\rm in}^{\rm pA}} \frac{\sigma_{\jpsi}}{\sigma_{\rm in}^{\rm NN}} \int \dd {\b} {\textstyle \binom{A}{N_{\rm p}}}  \left[ p_{\rm pA}(\b) \right]^{N_{\rm p}}  \left[ 1-p_{\rm pA}(\b) \right]^{A-N_{\rm p}}   \, .
\ee

The normalized probability distribution for the number of collisions of the projectile proton in the events tagged by both centrality and $\jpsi$ production is
\be
\label{eq:PNpartpAtag}
 P(N_{\rm p}|\jpsi,[N_1,N_2]) = \cfrac{  \tfrac{\sigma_{\jpsi}}{\sigma_{\rm in}^{\rm NN}} N_{\rm p} \int \dd{\b} \binom{A}{N_{\rm p}}  \left[p_{\rm pA}(\b)\right]^{N_{\rm p}}  \left[1-p_{\rm pA}(\b)\right]^{A-N_{\rm p}}}{\sum\limits_{N=N_1}^{N_2} \tfrac{\sigma_{\jpsi}}{\sigma_{\rm in}^{\rm NN}} N \int \dd{\b} \binom{A}{N}  \left[p_{\rm pA}(\b)\right]^{N}  \left[1-p_{\rm pA}(\b)\right]^{A-N}} 
\ee
and is independent of the hard process cross section. The corresponding average of $N_{\rm p}$ is
\be
\label{eq:NeffpA}
\left.\langle N_{\rm p} \rangle \right|_{\jpsi,[N_1,N_2]} = 1+\cfrac{\sum\limits_{N=N_1}^{N_2} N(N-1) \int \dd \b \binom{A}{N} \left[p_{\rm pA}(\b)\right]^{N}  \left[ 1-p_{\rm pA}(\b)\right]^{A-N}}{\sum\limits_{N=N_1}^{N_2}  N \int \dd{\b} \binom{A}{N} \left[ p_{\rm pA}(\b)\right]^{N}  \left[1-p_{\rm pA}(\b)\right]^{A-N}} \, .
\ee
The interpretation of \eq{eq:NeffpA} is straightforward. Unity stands for the target nucleon which participated to the hard process. The second term corresponds to the target nucleons which also undergo an inelastic collision with the projectile, and may thus contribute to the transverse momentum broadening of the $c\bar c$ pair, with the probability $\sigma_{broad}/\sigma_{\rm in}^{\rm NN}$.

The effective path length in the target nucleus for the centrality class $[N_1,N_2]$ thus reads 
\be
\label{eq:LefpA}
L_{\rm A} = L_{\rm p}+ \cfrac{\sum\limits_{N=N_1}^{N_2} N(N-1) \int \dd{\b} \binom{A}{N}  \left[p_{\rm pA}(\b)\right]^{N}  \left[1-p_{\rm pA}(\b)\right]^{A-N}}{\sigma_{\rm in}^{\rm NN} \, \rho_0 \sum\limits_{N=N_1}^{N_2}  N \int \dd{\b} \binom{A}{N}  \left[p_{\rm pA}(\b)\right]^{N}  \left[1-p_{\rm pA}(\b)\right]^{A-N}} \, , 
\ee
where $L_{\mathrm p}$ is the corresponding length in a proton target. For the minimum bias case doing the summations is trivial and one recovers the expression used in Ref.~\cite{Arleo:2012rs}. In the numerical applications we take $L_{\rm p} = 1.5 \, {\rm fm}$ and $\rho_0=0.17$~fm$^{-3}$.

\subsubsection*{d--A (RHIC case)}

Compared to the \pA\  case some complications arise because of the deuterium projectile. What is relevant for the broadening is the number of collisions $N_t$ suffered by the nucleon of the deuterium
participating to the hard process.
This number is not equal to the number of participants in a given event. 
In the binomial expansion of the first multiplier in the numerator of \eqref{eq:PNpartdA} 
\be
\left[ p_{\rm dA}({\bf b}, {\bf r})\right]^{N_{\rm p}} = {\textstyle \binom{N_{\rm p}}{N_t}} \left[p_{\rm pA}({\bf b}+{\bf r}/2)\right]^{N_t} \left[p_{\rm pA}({\bf b}-{\bf r}/2) - p_2({\bf b},{\bf r})\right]^{N_{\rm p}-N_t} \, , 
\ee
each term corresponds to the probability of $N_t$ collisions of the tagged nucleon of the deuterium projectile with overall $N_{\rm p}$ participants in the target nucleus. The joint probability of the production process in d--A collision with given  $N_t$  and overall $N_{\rm p}$ thus reads:
\begin{multline}
\label{eq:PhardNtNpA}
P(N_{\rm p},N_t,\jpsi) =  \cfrac{ 2 N_t}{\sigma_{\rm in}^{\rm dA}} \frac{\sigma_{\jpsi}}{\sigma_{\rm in}^{\rm NN}} \int \dd \b \int \dd \r  P_{\rm d}({\r}) \times \\
{ {\textstyle \binom{A}{N_{\rm p}}   \binom{N_{\rm p}}{N_t}} \left[p_{\rm pA}({\bf b}+{\bf r}/2)\right]^{N_t} \left[p_{\rm pA}({\bf b}-{\bf r}/2) - p_2({\bf b},{\bf r})\right]^{N_{\rm p}-N_t} \left[1- p_{\rm dA}(\b,\r) \right]^{A-N_{\rm p}}} \, .
\end{multline}

Correspondingly the number of inelastic rescatterings for the nucleon of the deuterium projectile tagged by the hard process is
\begin{multline}
\label{eq:NeffdA}
 \left.\langle N_t \rangle\right|_{\jpsi,[N_1,N_2]} = 1+ \\ \cfrac{\sum\limits_{N=N_1}^{N_2} N(N-1) \int \dd{\b} \int \dd \r  P_{\rm d}({\r}) \binom{A}{N}  [p_{\rm pA}(\b)]^2 \left[p_{\rm dA}(\b,\r)\right]^{N-2}  \left[1-p_{\rm dA}(\b,\r)\right]^{A-N}}
 {\sum\limits_{N=N_1}^{N_2} N(N-1) \int \dd{\b} \int \dd \r  P_{\rm d}({\r}) \binom{A}{N}  p_{\rm pA}(\b) \left[p_{\rm dA}(\b,\r)\right]^{N-1} \left[1-p_{\rm dA}(\b,\r)\right]^{A-N}}
\end{multline}
Interpretation of \eqref{eq:NeffdA} goes along the same line as for \eqref{eq:NeffpA}. 

The path length of the $c\bar c$ pair for a given centrality class is thus (after substituting $p_{\rm pA}(\b) = \sigma_{\rm in}^{\rm NN} \, T_{\rm A}(\b)$):
\be
\label{eq:LefdA}
 L_{\rm A} = L_{\rm p} + \cfrac{\sum\limits_{N=N_1}^{N_2} N(N-1) \int \dd{\b} \int \dd \r  P_{\rm d}({\r}) \binom{A}{N}  (T_{A}(\b))^2 \left[p_{\rm dA}(\b,\r)\right]^{N-2}  \left[1-p_{\rm dA}(\b,\r)\right]^{A-N}}{\rho_0 \sum\limits_{N=N_1}^{N_2} N(N-1) \int \dd{\b}\int \dd \r  P_{\rm d}({\r}) \binom{A}{N}  T_{A}(\b) \left[p_{\rm dA}(\b,\r)\right]^{N-1}  \left[1-p_{\rm dA}(\b,\r)\right]^{A-N}} 
\ee
For the minimum bias case ($N_1=1$, $N_2=A$) the summations in \eq{eq:LefdA} are explicit and one again recovers the expression (3.18) of Ref.~\cite{Arleo:2012rs}. The path length for the different centrality classes is given in Table~\ref{tab:centr-Npart}.

\bibliography{}

\ed